\documentclass[%
 reprint,
 amsmath,amssymb,
 aps,superscriptaddress
]{revtex4-2}

\usepackage{graphicx}
\usepackage{dcolumn}
\usepackage{bm}
\usepackage{verbatim}
\usepackage{subfigure}
\usepackage{multirow}
\usepackage{setspace}
\usepackage{physics}
\usepackage{xcolor}
\usepackage{siunitx}

\usepackage{amsmath,amsfonts,amssymb}
\usepackage[subfigure]{tocloft}
\usepackage{adjustbox}

\renewcommand{\vec}[1]{\ensuremath{\mathbf{#1}}}

\newcommand{\pes}{\mathcal{E}}

\newcommand{\conc}{\elec\!/\,\text{f.u.}}
\newcommand{\gpoint}{\boldsymbol{\Gamma}}

\newcommand{\elec}{{e^-}}

\newcommand{\bto}{BaTiO$_3$}

\newcommand{\pp}{\vec{P}}
\newcommand{\ppeq}{\bar{\vec{P}}}
\newcommand{\dip}{\vec{p}}

\newcommand{\dipa}{\vec{p}_{\alpha}}

\newcommand{\aveu}{\bar{u}}
\newcommand{\Eg}{{>\!\!E_g}}
\newcommand{\Deltatio}{\Delta_{\text{Ti-O}}}
\newcommand{\aveDeltatio}{\bar{\Delta}_{\text{Ti-O}}}

\newcommand{\ppfm}{\bar{\pp}_{\text{FM}}}
\newcommand{\pfm}{\pp_{\text{FM}}}
\newcommand{\Qfm}{Q_{\text{FM}}}
\newcommand{\aveQfm}{\bar{Q}_{\text{FM}}}
\newcommand{\bond}{d_{\text{Ti-0}}}

\newcommand{\bondcubic}{d_{\text{Ti-0}}^{\text{Pm$\bar{3}$m}}}

\begin{document}

\preprint{APS/123-QED}

\title{Ultrafast polarization switching in BaTiO$_3$ by photoactivation of its ferroelectric and central modes}
\author{Fangyuan Gu}
\affiliation{%
 Tsung-Dao Lee Institute, Shanghai Jiao Tong University, Pudong, Shanghai 201210, China}
\email{fangyuan.gu@sjtu.edu.cn}

\author{Paul Tangney}%
\affiliation{%
Department of Physics and Department of Materials, Imperial College London, London SW7 2AZ, UK}

\date{\today}

\begin{abstract}
We use molecular dynamics simulations with machine-learned atomistic force fields
to simulate photoexcitation of \bto~by a femtosecond laser pulse whose photon energy
exceeds the optical gap. We demonstrate selective displacive
excitation of coherent zone-center ferroelectric mode phonons and of 
the strongly anharmonic central mode. We show that
the direction of $\pp$ can either be reversed by a pulse in
hundreds of femtoseconds
or, on a longer time scale and when combined with a weak field, switched to 
any one of its symmetry-equivalent directions.
\end{abstract}

\maketitle

The demand for faster and more efficient optoelectronic devices has motivated a lot
of research into the use of femtosecond (fs) laser pulses to quickly switch polarization ($\pp$) domains
in ferroelectric perovskites~\cite{Fahy_1994, nelson_1996, Qi_2009, subedi_2015, chen_2016, mankowsky_2017, freeland_2018, garcia_2018, gruverman_2018, Peng_2022, Bellaiche_PRB_2023}.
\bto~is a widely used and intensively studied ferroelectric material, which is often regarded
as a prototypical ferroelastoelectric perovskite.
Therefore understanding its interaction with ultrashort laser pulses is both fundamentally 
interesting and of practical importance to device design and innovation:
The speed with which $\pp$ and other ferroic orders can reliably be manipulated 
is steadily reducing as experimental methods and device designs are gradually refined, 
and as new manipulation mechanisms are discovered, such as the one presented in this Letter~\cite{cavalleri_2006,plech_2007,Qi_2009,fernandez_2015,chen_2016,mankowsky_2017,freeland_2018,garcia_2018,gruverman_2018,Bagri_2022}.

We use atomistic molecular dynamics (MD)
simulations of ferroelectric (FE) \bto~ to demonstrate
that above-optical-gap (${\Eg}$) photoexcitation 
with a fs laser pulse can reverse the direction of
$\pp$ within ${\sim 100\;\text{fs}}$,  or
lower the coercive field strength ($E_c$) for long
enough to switch it with a relatively-weak applied 
field. It can also induce a temporary displacive transition 
to the unpolarized cubic Pm$\bar{3}$m structure of 
\bto's paraelectric (PE) phase. This structure would spontaneously
polarize again, via a quasi-random process of domain nucleation
and growth, when the photoexcited carriers recombined or dispersed.
By biasing this process with an applied field or GHz/THz pulses, 
$\pp$ could be manipulated into any one of its symmetry-equivalent directions.

As in previous works~\cite{Tangney_1999,Tangney_2002,Murray_2005PRB, Garcia_PRB_2006, Murray_PRB_2007, fahy_science_2007, Gu2021,Bellaiche_PRB_2023}, 
we approximate the absorption of a fs ${\Eg}$ pulse as an instantaneous change 
to the state of the electrons, which takes them out of thermal equilibrium with the lattice, 
and creates two separate thermalized populations of carriers: conduction band electrons and valence band holes. 
These carriers' densities ($x$) are equal, initially, and remain approximately constant
for several ps~\cite{Tangney_1999,Tangney_2002,Murray_2005PRB,Murray_PRB_2007, fahy_science_2007, Gu2021,Bellaiche_PRB_2023}.

Although we neglect the ${\sim 100}$'s of fs~\cite{Sundaram2002,Gamaly_2011,shah_2013,Phillips_2015} 
taken for the populations of electrons and holes to thermalize, 
the physical mechanism by which a fs ${\Eg}$ pulse interacts with
$\pp$ does not require this thermalization, or wait for it to happen. 
It is known as \emph{displacive excitation of coherent phonons} (DECP)~\cite{zeiger1992, Hunsche_PRL1995, Tangney_1999, bargheer2004, Murray_2005PRB, Tangney_2002, Gu2021}
and
it begins as soon as electrons vacate bonding states and occupy
anti-bonding states, because it is driven by the forces on the crystal's sublattices caused by this 
change of the electron density. Qualitatively, and semi-quantitatively, these forces are determined by $x$ and by
the characters of the upper valence band states ($\approx$ {O-2p} admixed with {Ti-3d}) and 
lower conduction band states ($\approx$ {Ti-3d} admixed with {O-2p}). They are relatively
insensitive to how holes and electrons, respectively, are distributed among these states~\cite{Gu2021}.

DECP occurs when a high density of photoexcited carriers is created by a fs laser pulse in 
a crystal that possesses A$_1$ phonon modes.
A$_1$ phonons are excited by $\Eg$ photoexcitation 
because the meaning of a mode having A$_1$ symmetry is that both its equilibrium and average
mode coordinates are not constrained by symmetry. Therefore they are changed, to some degree, by any stimulus.
When a laser pulse changes a crystal's A$_1$ mode coordinates
suddenly, by redistributing electron density and weakening bonds, 
the crystal's sublattices suddenly have the wrong relative displacements.
Therefore they move along the A$_1$ eigenvectors
towards the new A$_1$ coordinates, which they overshoot and 
oscillate about~\cite{Hunsche_PRL1995,Tangney_1999,Gu2021}.
This oscillation is the displacively-excited coherent A$_1$ phonon.

\bto~has three FE phases, which all possess A$_1$ modes and 
have almost identical electronic structures.
Each FE phase only differs from the Pm$\bar{3}$m structure
by tiny symmetry-breaking relative displacements of its sublattices
along its A$_1$ eigenvectors, which lower the 
potential energy by ${\Delta U\equiv U_\text{Pm$\bar{3}$m}-U_\text{FE}>0}$,
and create a $\pp$ field~\cite{Gu2021}.
By far the largest contributions to both $\pp$ and ${\Delta U}$ 
come from the \emph{polar distortion} of Pm$\bar{3}$m along the eigenvector
of the FE phase's A$_1$ \emph{ferroelectric mode} (FM), which 
is a counter-motion of the Ti and O sublattices along an
axis parallel to $\pp$. 
The polar disortion makes the Ti-O Coulombic attraction
more negative by shortening the Ti-O bond length, and
the displacements along the other A$_1$ eigenvectors 
help to accommodate it~\cite{Gu2021}. 

We simulated ultrafast $\Eg$ photoexcitation of  
\bto's R3m FE phase, which has three optical A$_1$ modes; namely, 
the FM, the \emph{Axe mode} (AM)~\cite{axe_1967}, and the \emph{Last mode} (LM)~\cite{last_1957}.
Both the FM and its counterpart in Pm$\bar{3}$m, which does not have A$_1$ symmetry,
are often referred to as the \emph{soft mode} or the \emph{Slater mode} (SM)~\cite{slater_1950}.
We refer to it as the FM when its A$_1$ symmetry is relevant and as the SM otherwise.
Ultrafast ${\Eg}$ photoexcitation induces motion along every A$_1$ eigenvector
to some degree, but it \emph{selectively} excites motion along the FM eigenvector
in the sense that the AM and LM are excited to much lesser degrees.
Before demonstrating this selectivity, we briefly explain it. We discuss it
in greater detail in Ref.~\onlinecite{Gu2021}.

Roughly-speaking, the SM of a given phase can be viewed as an oscillation of 
${\Deltatio\equiv\bondcubic-\bond\geq 0}$, where $\bond$ and $\bondcubic$ are the 
{Ti-O} nearest-neighbour distances in the given phase and in Pm$\bar{3}$m, respectively.
We choose the origin for the FM mode coordinate, $\Qfm$, to be 
where the polar distortion vanishes, i.e., in the Pm${\bar{3}}$m phase. 
Therefore the thermodynamic averages of  $\Qfm$, $\Deltatio$, $\pp$, 
and the contribution, $\pfm$, of the polar distortion to $\pp$,
approximately satisfy
${\ppeq(T,x)\approx \ppfm(T,x)\propto\aveQfm(T,x)\propto\aveDeltatio(T,x)}$.
Photoexcited carriers weaken the {Ti-O} attraction by screening it and by
reducing the magnitudes of Ti and O ions' charges~\cite{Gu2021}. 
They reduce charges because promoting electrons from predominantly O-2p bonding states
to predominantly Ti-3d anti-bonding states moves some electron density from O to Ti.
Therefore DECP excites the FM strongly because ${\Qfm\propto\Deltatio}$ is highly sensitive to $x$. 
However, the AM and LM do not depend linearly on $\Deltatio$ 
and there is no obvious reason why DECP would excite them strongly.

We performed MD simulations with dipole-polarizable and variable-charge machine-learned atomistic
force fields, as described in the Supplementary Material and Ref.~\onlinecite{nemytov_thesis}.
We parameterized three force fields:
To model interactions before absorption of a laser pulse we fit the parameters
to density functional theory (DFT) calculations of thermally-disordered crystals
with electrons in their ground state (${x=0}$). 
To model interactions after pulse absorption, we parameterized
force fields for 
${x=0.05}$ electrons per \bto~formula unit ($\conc$) and  ${x=0.12\;\conc}$ by
fitting the parameters 
to constrained-DFT calculations,
as described in Refs.~\onlinecite{Tangney_1999,Tangney_2002, Gu2021}.
We used a ${12\times 12\times 12}$ supercell ($8640$ atoms), under
periodic boundary conditions, and performed
long MD simulations with the ${x=0}$ potential to equilibrate, before 
modelling fs ${\Eg}$ pulse absorption by switching abruptly to 
one of the photoexcited potentials.
We calculated the $\pp$
autocorrelation function, ${\langle{\mathbf{P}(t_0)\mathbf{P}(t_0+t)}\rangle_{t_0}}$,
from the first ${10\;\text{ps}}$ after photoexcitation
and Fourier transformed it to calculate the infrared (IR) absorption spectrum.

Both ${\Delta U}$ and the FE to PE transition temperature, $T_C$,
are highly sensitive to strain and are lowered by compression~\cite{goldschmidt_ratio}.
Therefore, when force fields or DFT overestimate the density, it
is common to perform calculations at the experimental density or
under negative pressure~\cite{zhong_PRL_1994,fallon_thesis}.
We found  ${T_C\approx 150~\text{K}}$, ${T_C\approx 100~\text{K}}$, 
and ${T_C\approx 50~\text{K}}$ 
for our ${x=0}$, ${x=0.05\;\conc}$, and ${x=0.12\;\conc}$ force fields, respectively.
However we chose not to apply negative pressure because working at a low $T$ 
allowed us to observe the DECP mechanism with less thermal noise,
and to calculate spectra with signal-to-noise ratios closer to those
that would be obtained with simulation cells comparable in size
to the photoexcited regions in pump-probe experiments.

\begin{figure}[!ht]
\centering     
\includegraphics[width=8.0cm]{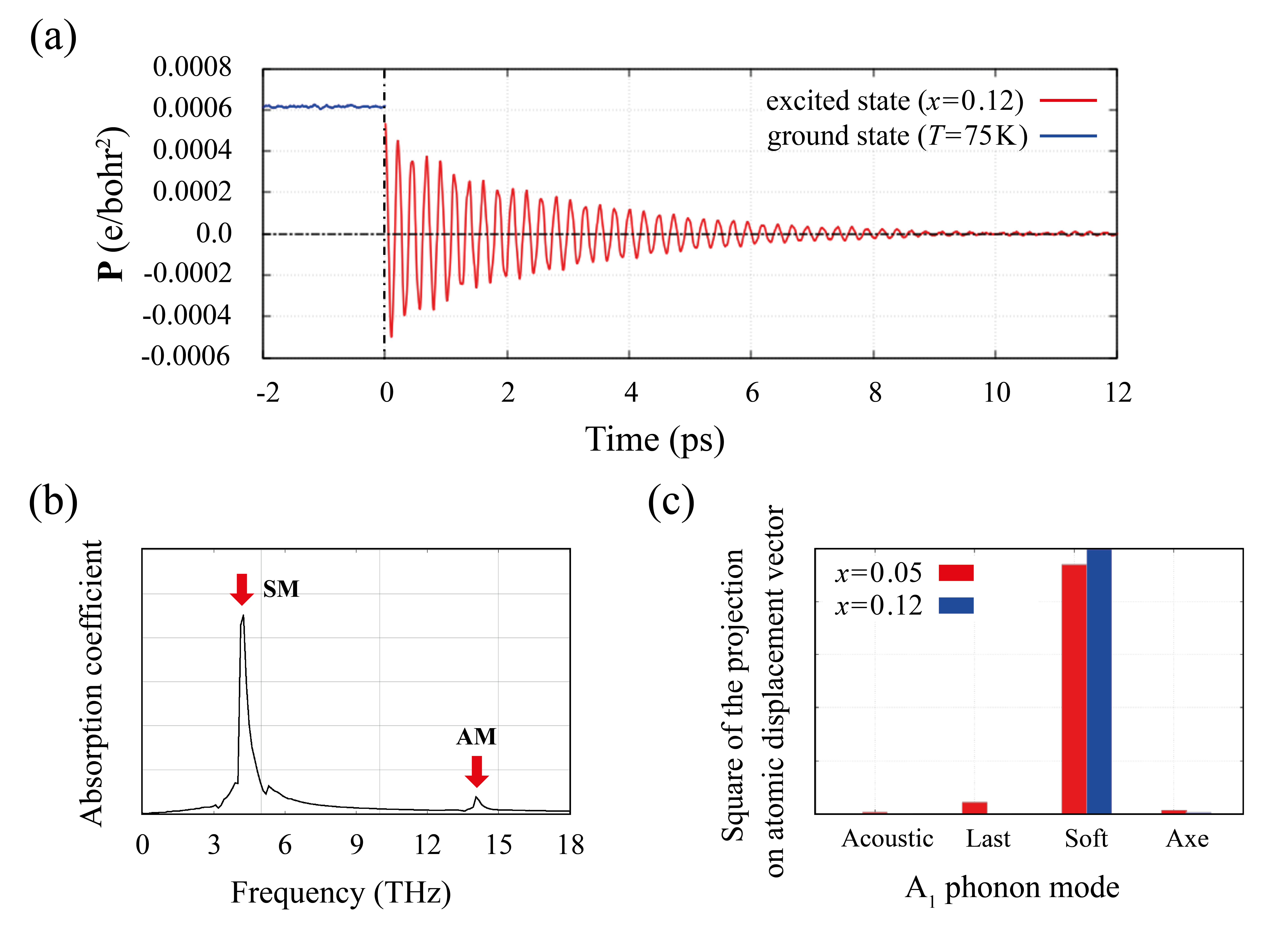}
\caption{(a) Polarization, $\vec{P}$, as a function of time ($t$), with ${\Eg}$ photoexcitation to a carrier
density of ${x=0.12\;\conc}$ occurring at ${t=0}$;
(b) The IR absorption spectrum immediately after photoexcitation;
(c) Squared projections of the ${\sqrt{\text{mass}}}$-scaled atomic displacement vectors onto the three
zone-center A$_1$ optical phonon eigenvectors. The sums of the squared projections are one.
}
\label{fig:selective_decp}
\end{figure}
Figure~\ref{fig:selective_decp}(a) is a plot of ${\pp}$ as function of time ($t$)
in MD simulations of photoexcitation to a carrier density of ${x=0.12\;\conc}$ at ${T=75\;\text{K}}$. 
At ${t=0}$, photoexcitation changes the value of $\pp$ at thermal equilibrium 
from ${\ppeq(75\;\text{K},0)\approx 6\times10^{-4}\;e^-/\text{bohr}}$ to
${\ppeq(75\;\text{K},0.12\;\conc)=0}$. The latter vanishes because
Pm${\bar{3}}$m is the thermodynamically stable phase at ${(T,x)=(75\;\text{K},0.12\;\conc)}$
with our force field.
Therefore the fs pulse causes $\ppeq$ to vanish suddenly as a consequence of ${\aveQfm}$ and $\aveDeltatio$ vanishing suddenly.
The change of $\aveQfm$ excites a large amplitude SM phonon by displacively exciting motion along the FM eigenvector.
This manifests in Fig.~\ref{fig:selective_decp} as a 
damped oscillation of $\pp$ about ${\pp=0}$, with an initial amplitude of ${\abs{\ppeq(75\;\text{K},0)}}$.
Figure~\ref{fig:selective_decp}(b) is the IR absorption
spectrum calculated immediately after photoexcitation, 
and Fig.~\ref{fig:selective_decp}(c) shows the decomposition,
into components along the A$_1$ eigenvectors, of the lattice's displacement 
from its new equilibrium immediately after photoexcitation.
These plots demonstrate that DECP selectively excites the SM.

To better understand
what happens when a fs $\Eg$ pulse is absorbed, it is useful
to regard the FM as an oscillation of $\pp$.
If $\dipa$ denotes the dipole moment of the ${\alpha^\text{th}}$ primitive cell of the crystal
divided by its volume, then $\pp$ is the average of ${\dipa}$ 
over all cells $\alpha$; and ${\ppeq(T,x)}$ is the value shared by $\pp$ and the time-average
of each ${\dipa(t)}$ at thermal equilibrium. 
Therefore a displacively-excited FM phonon can be viewed
as a collective motion of the set ${\{\dipa\}_\alpha}$ of all ${\dip}$'s, which is caused
by a sudden change of $\ppeq$ from  ${\ppeq(T,0)}$ to ${\ppeq(T,x)}$, 
and which has an initial amplitude of ${\Delta\ppeq(T,x)\equiv \abs{\ppeq(T,0)-\ppeq(T,x)}}$.
The motion is collective in the statistical sense that
the \emph{average} time derivative of the $\dip$'s is
finite and remains finite until the crystal reaches
a new thermal equilibrium in which the time averages
of the ${\dip}$'s are all equal to ${\ppeq(T,x)}$.

\begin{figure}
\includegraphics[width=8.5cm]{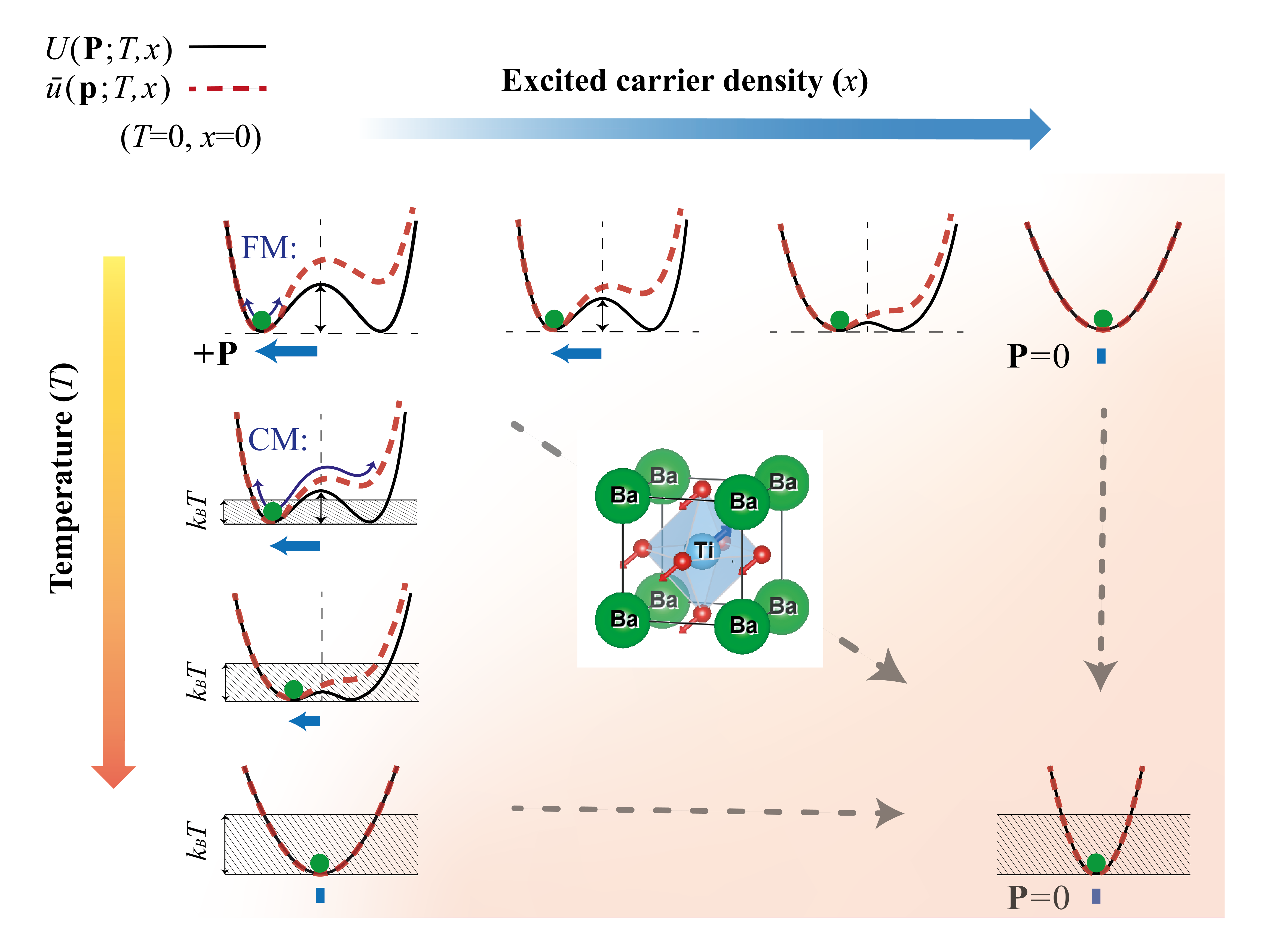}
\centering
\caption{
Schematic illustrating the average potential energy, $U$ and $\aveu$,
as a function of $\pp$ and $\dip$, respectively, at ${T=0}$ in the electronic ground state and at several 
temperatures ($T$) and photoexcited carrier densities ($x$). The BaTiO$_3$ crystal structure in the center demonstrates the FM eigenvector of the R3m phase.
}
\label{fig:double_well}
\end{figure}

Now consider a simple model of the crystal in which ${\ppeq=\ppfm}$, $\dipa$ completely specifies
the structure of the ${\alpha^\text{th}}$ cell,
and ${u_\alpha(\dipa;T,x,t)}$ denotes the potential energy of the entire crystal, as a function
of $\dipa$, when all other ${\dip}$'s are fixed at their values at time $t$.
Let ${\aveu(\dip;T,x)}$ denote
the average of $u_\alpha$, over all $\alpha$ or over time, at thermal equilibrium; and
let ${U(\pp;T,x)}$ denote the thermodynamic 
average of the potential energy over all microstates of the crystal 
for which ${\pfm=\pp}$. 
Each $u_\alpha$ is time dependent because it
is highly sensitive to the structures and strains of surrounding cells~\cite{fallon_thesis}.
Instantaneously, it is not symmetric about ${\dipa=0}$, and it may be a single well or an
asymmetric double well, with the (deeper) minimum continuously moving as the $\dip$'s
of surrounding cells change~\cite{fallon_thesis}.
However ${U(\pp;T,x)}$ and ${\aveu(\dip;T,x)}$ are
independent of $t$ because they are thermodynamic averages. 
Figure~\ref{fig:double_well} shows schematic cross sections of them, along 
the axis passing through $\ppfm$ and ${-\ppfm}$, as both $T$ and $x$
are varied. When ${(T,x)\approx(0,0)}$, $U$ is a symmetric double well, with the wells
at ${\pp=\pm\ppfm(T,x)}$ corresponding to symmetry-equivalent R3m structures.
The energy barrier separating them is at ${\pp=0}$, which
corresponds to the Pm$\bar{3}$m structure within this simple model, 
and its height is ${\Delta U(T,x)\equiv U(0;T,x)-U(\ppfm(T,x);T,x)}$.

\begin{figure}[!ht]
\centering     
\includegraphics[width=8.3cm]{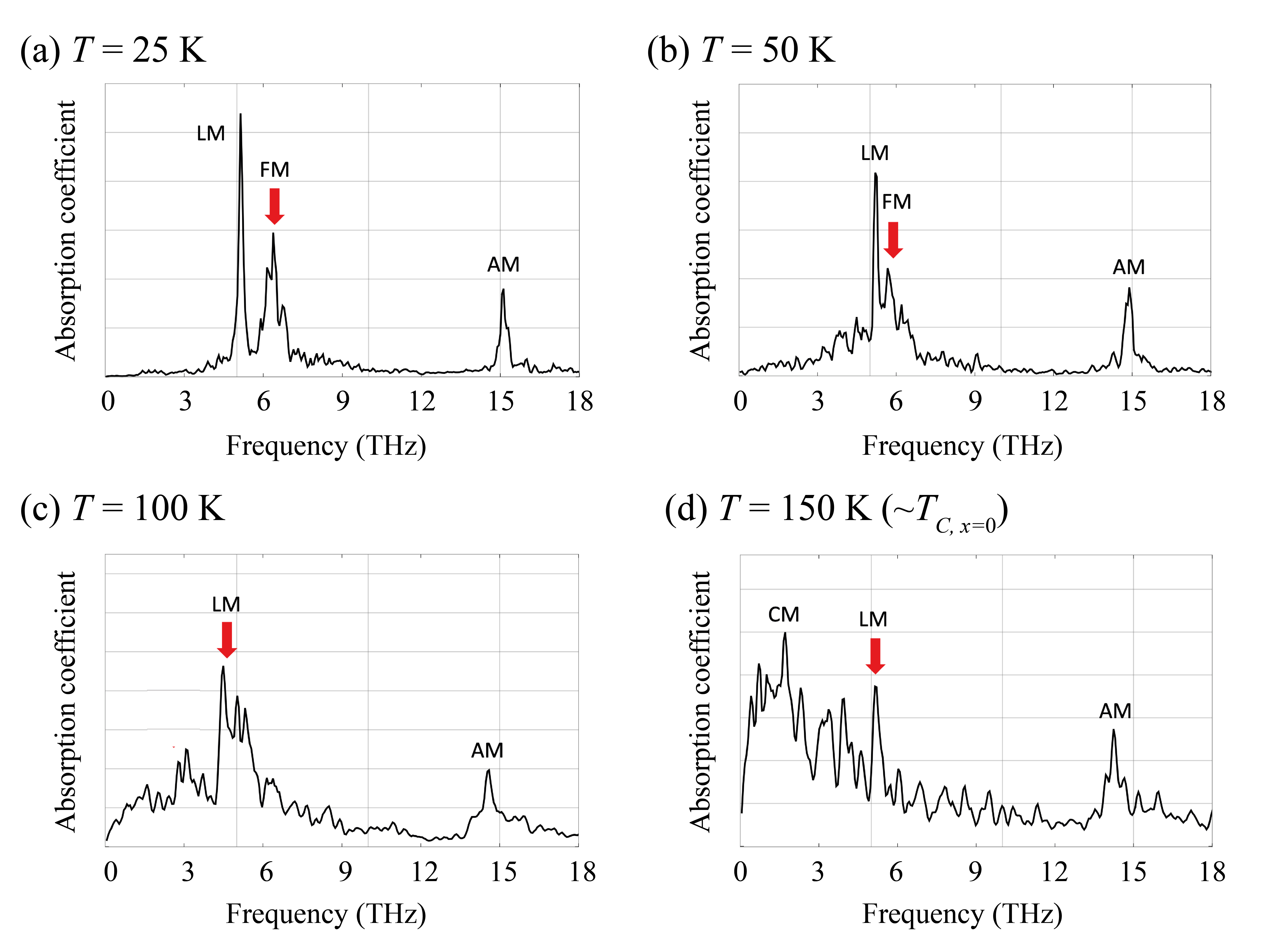}
\caption{
Infrared absorption spectra in the electronic ground state. 
}
\label{fig:infrared}
\end{figure}

The zone-center FM is a coherent collective oscillation of the $\dip$'s about
${\dip=\ppfm}$. 
Both experimentally~\cite{bellaiche_2008,weerasinghe_bellaiche_2013}, 
and in our ${x=0}$ simulations (Fig.~\ref{fig:infrared}), increasing
$T$ causes the FM's 
IR absorption peak to soften and broaden, and a 
very broad peak, known as the \emph{central mode} (CM), to emerge in the frequency range ${0-3\;\text{THz}}$.
The CM is not one of the crystal's normal modes, and it does not exist in the ${T\to 0}$ limit.
It gradually becomes active as $T$ increases and the directions of the $\dip$'s become disordered.

It is common to view the dynamics of each ${\dipa}$ as motion on
a potential energy surface with eight minima~\cite{blinc_2004, Senn_PRL_2016, Gu2021, ceriotti_2022}. At each minimum, 
${\dipa}$ is parallel to one of the four body diagonals of the
cubic cell shown in Fig.~\ref{fig:double_well},
and is directed towards a different one of the eight corner Ba atoms~\cite{bersuker_1966,comes_1968,chaves_1976}.
The CM is often thought of as a collective hopping motion of the $\dip$'s between
two or more of these eight minima. However, Fallon's 
calculations of ${u_\alpha}$ for various structures of surrounding cells
suggest that this picture may be simplistic
(Ref.~\onlinecite{fallon_thesis}, Sec.~7.4). 
Rather than $\dipa$ hopping
between eight ever-present minima of a relatively-passive potential energy surface, 
it may move on a surface with only one minimum. The minimum might \emph{lead}
its dynamics by moving rapidly as the $\dip$'s of surrounding cells change.
Therefore in a more realistic eight-site hopping model, it might be the
minimum of $u_\alpha$ that hops between sites, while $\dipa$ simply follows it.

Regardless of how active a role the time dependence of $u_\alpha$ plays, 
the CM peak is the spectral signature of the relatively slow and 
anharmonic large-amplitude `rattling' of the ${\dip}$'s 
between multiple directions, which emerges as they gain enough
thermal energy to change direction.
At low $T$, when most $\dip$'s are aligned, 
$\dipa$ spends most of its time near the ${\dipa\parallel\pp}$ site. 
As $T$ increases 
it spends an increasing fraction of its time at the other seven sites. Therefore
the directional disorder of the $\dip$'s reduces
$\abs{\ppeq}$ and $\abs{\ppfm}$ and, if our revision of the eight-site model is realistic, 
it makes $\aveu$ more symmetric because the minimum of $u_\alpha$ spends more of its
time at the ${\dipa\parallel(-\pp)}$ site.
Disorder also reduces ${\Delta U}$
because the potential energy is lower when each ${\dip}$ is parallel to its neighbours. 
Reducing ${\Delta U/(k_B T)}$ increases
the proportion of time for which the direction of each $\dip$ differs significantly from that of $\pp$,
and reduces the fraction of the $\dip$'s that, at any given time, 
are participating in the FM, i.e., 
performing small synchronized oscillations 
about energy minima at their ${\dip\parallel\pp}$ sites. 
Therefore, when the CM becomes active it amplifies itself by generating disorder
that makes it easier for the $\dip$'s to change direction.

The FM IR absorption peak shrinks as the CM peak grows with increasing $T$
because, as more $\dip$'s contribute to the CM, fewer are available to participate in it.
It also softens and broadens
because reducing  ${\Delta U}$ makes the wells in $U$ shallower, which reduces
their curvatures and makes them less harmonic.
As $T$ increases even further, the $\dip$'s becomes so
disordered that ${\Delta  U}$ vanishes and
$U$ becomes a single well with a minimum at ${\pp=0}$. 
At the lowest values of 
$T$ for which Pm$\bar{3}$m is stable, $U$ is approximately quartic (i.e., flat-bottomed; see Fig.~\ref{fig:double_well}),
meaning that a sufficiently-small polar distortion neither raises nor lowers $U$.
When $T$ is larger,  $U$ is quadratic near its minimum and its curvature increases as $T$ increases~\cite{Gu2021}.

The effects on $\aveu$ and $U$ of increasing $x$ are similar
to the effects of increasing $T$:  by weakening the Ti-O attraction, photoexcited carriers 
reduce both ${\Delta U}$ and the magnitude of the polar distortion~\cite{Gu2021}. 
Therefore increasing $x$ reduces ${\ppfm}$ by moving the two minima of $U$ closer together and,
by making the two energy wells shallower~\cite{Gu2021}, it
lowers the FM frequency, makes it less harmonic, and makes the CM more active.
Therefore it decreases the proportion of time for which 
each ${\dip}$ is approximately parallel to $\pp$. 

There is no CM peak in Fig.~\ref{fig:selective_decp}(b) because, at ${(T,x)=(75\;\text{K},0.12\;\conc)}$, 
the combined effects of $x$ and $T$ make $U$ a single approximately-quadratic well. 
Instead of the $\dip$'s rattling between different directions with very large amplitudes, as they would at lower 
values of $x$ or $T$, their collective motion is a superposition of the Pm$\bar{3}$m phase's 
three degenerate SMs. When $x$ is large or ${T-T_C}$ is large and positive, each zone-center 
SM of Pm${\bar{3}}$m is an oscillation of one of three
mutually-orthogonal components of $\pp$ about the approximately-quadratic minimum
of a function identical to ${U(\pp;T,x)}$.
When ${x}$ and/or ${T}$ are either very large or very small, the CM is not active, 
the minima of $U$ and $\aveu$ have relatively-high curvatures, and thermal fluctuations of the
${\dip}$'s are much smaller than when the CM is active.

\begin{figure}
\centering
\vspace{10pt}
\includegraphics[width=8.3cm]{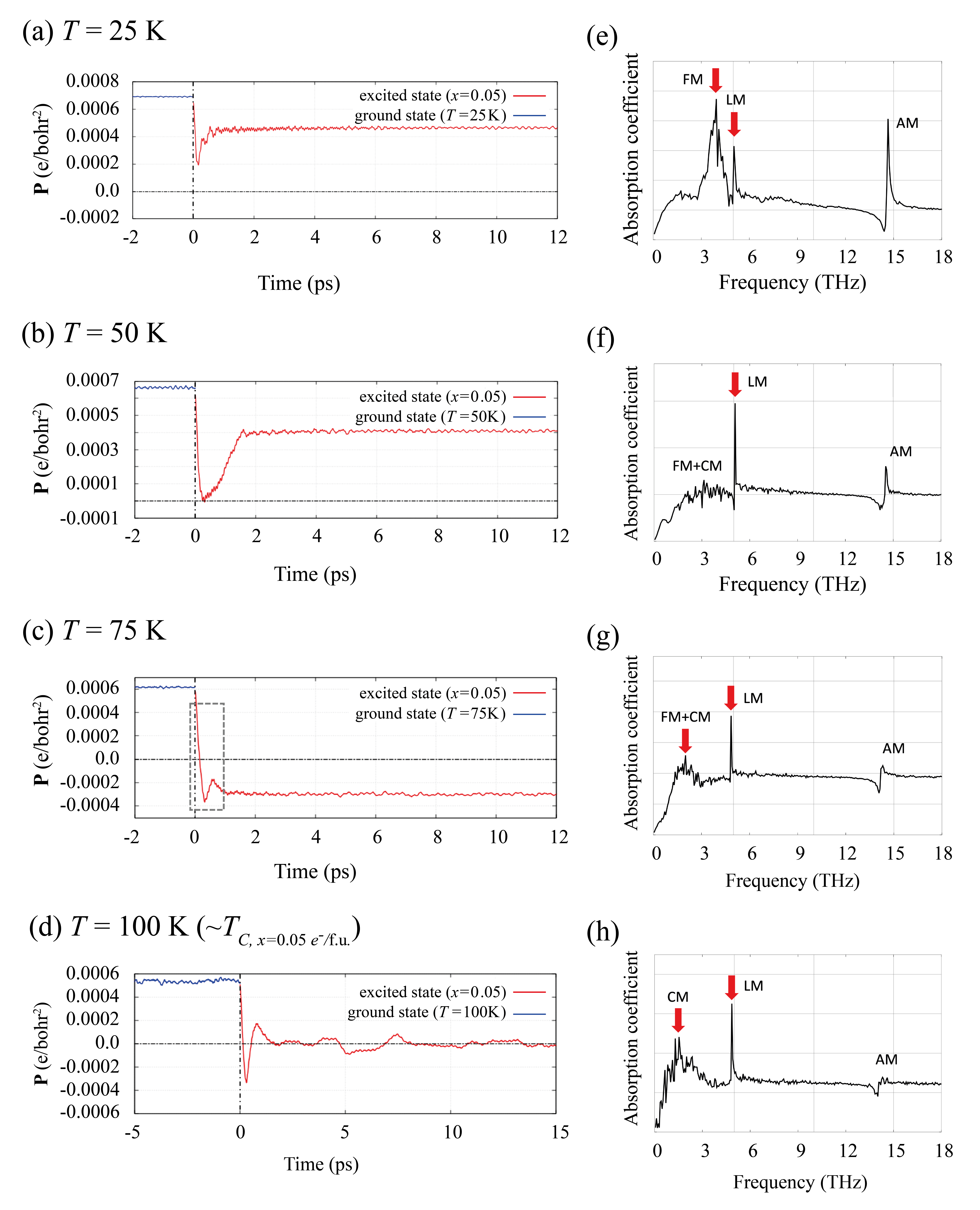}
\caption{
(a)-(d): $\vec{P}$ as a function of time in $NPH$ simulations
of ultrafast photoexcitation to a carrier density of
of ${x=0.05\;\conc}$ at four temperatures.
(e)-(h): IR absorption spectra at the same four temperatures immediately after the simulated
pulse absorption at ${t=0}$.
}
\label{fig:Pcontrol}
\end{figure}
Figures~\ref{fig:Pcontrol}(a)-(d)
are plots of ${\pp(t)}$ in simulations 
of photoexcitation to a much smaller carrier density (${x=0.05\;\conc}$) than
in the simulations reported in Fig.~\ref{fig:selective_decp}. 
R3m is stable at
${(T,x)=(75\;\text{K},0.05\;\conc)}$, but 
Pm$\bar{3}$m is stable at ${(T,x)=(100\;\text{K},0.05\;\conc)}$. 
The IR spectra in Figs.~\ref{fig:Pcontrol}(e)-(h)
show the emergence of the CM as $T$ increases, and that
the FM has softened before its peak disappears.
They also show that, at ${T=100\;\text{K}}$, the CM still has a substantial peak.
This implies that the oscillation of ${\pp}$ about ${\pp=0}$ in Fig.~\ref{fig:Pcontrol}(d)
is not simply a superposition of small-amplitude harmonic SMs.
It implies that the average magnitude of the $\dip$'s is large, 
that the $\dip$'s are disordered, that each $\dip$ is rattling between multiple directions with a large amplitude, 
and that $U$ and $\aveu$ are either flat, or shallow double wells.
This explains why the oscillations about ${\pp=0}$ in Fig.~\ref{fig:Pcontrol}~(d)
are so much less harmonic than those in Fig.~\ref{fig:selective_decp}~(a), 
and why, when ${T\leq 75\;\text{K}}$, 
the damping of the displacively-excited motion along the FM eigenvector 
is strong enough for $\pp$ to stabilize at ${\pp\approx\ppeq(T,0.05\;\conc)}$
almost immediately. 

This damping of the collective component of the motion of the $\dip$'s,
which can be viewed as their motions falling out of synchronicity,
is a crucial ingredient in the pulse-induced $\pp$-reversal mechanism that we propose, 
and which Fig.~\ref{fig:Pcontrol}(c) demonstrates. Without it,
$\pp$ would return to its original direction almost as quickly 
as it reversed. In Fig.~\ref{fig:Pcontrol}(c), 
$\pp$ reverses in less than half a FM period and remains reversed.
This demonstrates that, when $T$ is low enough that the FE phase is stable,
there exists a pulse fluence for which $\pp$ is
deterministically and permanently reversed within $100$'s of fs
of pulse absorption.

Pulse-induced $\pp$-reversal is permanent, but the
reductions of $E_c$ and $\abs{\ppeq}$, 
and the photoinduced stability of Pm$\bar{3}$m (Fig.~\ref{fig:selective_decp}(a) and Fig.~\ref{fig:Pcontrol}(d)),
only last until $x$ is reduced by electron-hole recombination and/or
diffusion.
During this time, which might be as short
as tens of ps or as long as many nanoseconds~\cite{Sundaram2002}, 
the ultimate direction of $\pp$ could be determined by a weak bias, such as an applied field, 
or by a different pulse-induced mechanism~\cite{Subedi_PRB_2015, mankowsky_2017, Peng_2022, Gayathri_2023}.

In summary, we have shown that fs $\Eg$ pulses would selectively excite motion 
along BaTiO$_3$'s A$_1$ FM eigenvector.
For a $T$-dependent range of pulse fluences, this motion would reverse $\pp$ 
within $100$'s of fs without subsequently returning it to its original direction.
Higher pulse fluences would induce a transient transition to the PE phase.
Therefore, a slower method of manipulating $\pp$, but one capable of placing it into
any of its symmetry-equivalent directions, would be to bias 
the process by which the transient PE phase
spontaneously repolarizes when electrons return to their ground state.

\begin{acknowledgments}
We acknowledge helpful discussions with \'Eamonn Murray, Wei Ku, Wei Wang, Chi-Ming Yim, and financial
support from
the International Postdoctoral Exchange Fellowship Program (YJ20210137) by the Office of China Postdoc Council (OCPC).
\end{acknowledgments}

\appendix

\bibliographystyle{apsrev4-2}
\bibliography{biblio}

\begin{thebibliography}{59}%
\makeatletter
\providecommand \@ifxundefined [1]{%
 \@ifx{#1\undefined}
}%
\providecommand \@ifnum [1]{%
 \ifnum #1\expandafter \@firstoftwo
 \else \expandafter \@secondoftwo
 \fi
}%
\providecommand \@ifx [1]{%
 \ifx #1\expandafter \@firstoftwo
 \else \expandafter \@secondoftwo
 \fi
}%
\providecommand \natexlab [1]{#1}%
\providecommand \enquote  [1]{``#1''}%
\providecommand \bibnamefont  [1]{#1}%
\providecommand \bibfnamefont [1]{#1}%
\providecommand \citenamefont [1]{#1}%
\providecommand \href@noop [0]{\@secondoftwo}%
\providecommand \href [0]{\begingroup \@sanitize@url \@href}%
\providecommand \@href[1]{\@@startlink{#1}\@@href}%
\providecommand \@@href[1]{\endgroup#1\@@endlink}%
\providecommand \@sanitize@url [0]{\catcode `\\12\catcode `\$12\catcode
  `\&12\catcode `\#12\catcode `\^12\catcode `\_12\catcode `\%12\relax}%
\providecommand \@@startlink[1]{}%
\providecommand \@@endlink[0]{}%
\providecommand \url  [0]{\begingroup\@sanitize@url \@url }%
\providecommand \@url [1]{\endgroup\@href {#1}{\urlprefix }}%
\providecommand \urlprefix  [0]{URL }%
\providecommand \Eprint [0]{\href }%
\providecommand \doibase [0]{https://doi.org/}%
\providecommand \selectlanguage [0]{\@gobble}%
\providecommand \bibinfo  [0]{\@secondoftwo}%
\providecommand \bibfield  [0]{\@secondoftwo}%
\providecommand \translation [1]{[#1]}%
\providecommand \BibitemOpen [0]{}%
\providecommand \bibitemStop [0]{}%
\providecommand \bibitemNoStop [0]{.\EOS\space}%
\providecommand \EOS [0]{\spacefactor3000\relax}%
\providecommand \BibitemShut  [1]{\csname bibitem#1\endcsname}%
\let\auto@bib@innerbib\@empty
\bibitem [{\citenamefont {Fahy}\ and\ \citenamefont
  {Merlin}(1994)}]{Fahy_1994}%
  \BibitemOpen
  \bibfield  {author} {\bibinfo {author} {\bibfnamefont {S.}~\bibnamefont
  {Fahy}}\ and\ \bibinfo {author} {\bibfnamefont {R.}~\bibnamefont {Merlin}},\
  }\href {https://doi.org/10.1103/PhysRevLett.73.1122} {\bibfield  {journal}
  {\bibinfo  {journal} {Phys. Rev. Lett.}\ }\textbf {\bibinfo {volume} {73}},\
  \bibinfo {pages} {1122} (\bibinfo {year} {1994})}\BibitemShut {NoStop}%
\bibitem [{\citenamefont {Wefers}\ \emph {et~al.}(1996)\citenamefont {Wefers},
  \citenamefont {Kawashima},\ and\ \citenamefont {Nelson}}]{nelson_1996}%
  \BibitemOpen
  \bibfield  {author} {\bibinfo {author} {\bibfnamefont {M.~W.}\ \bibnamefont
  {Wefers}}, \bibinfo {author} {\bibfnamefont {H.}~\bibnamefont {Kawashima}},\
  and\ \bibinfo {author} {\bibfnamefont {K.~A.}\ \bibnamefont {Nelson}},\
  }\href {https://doi.org/https://doi.org/10.1016/0022-3697(96)00008-X}
  {\bibfield  {journal} {\bibinfo  {journal} {J. Phys. Chem. Sol.}\ }\textbf
  {\bibinfo {volume} {57}},\ \bibinfo {pages} {1425 } (\bibinfo {year}
  {1996})}\BibitemShut {NoStop}%
\bibitem [{\citenamefont {Qi}\ \emph {et~al.}(2009)\citenamefont {Qi},
  \citenamefont {Shin}, \citenamefont {Yeh}, \citenamefont {Nelson},\ and\
  \citenamefont {Rappe}}]{Qi_2009}%
  \BibitemOpen
  \bibfield  {author} {\bibinfo {author} {\bibfnamefont {T.}~\bibnamefont
  {Qi}}, \bibinfo {author} {\bibfnamefont {Y.-H.}\ \bibnamefont {Shin}},
  \bibinfo {author} {\bibfnamefont {K.-L.}\ \bibnamefont {Yeh}}, \bibinfo
  {author} {\bibfnamefont {K.~A.}\ \bibnamefont {Nelson}},\ and\ \bibinfo
  {author} {\bibfnamefont {A.~M.}\ \bibnamefont {Rappe}},\ }\href
  {https://doi.org/10.1103/PhysRevLett.102.247603} {\bibfield  {journal}
  {\bibinfo  {journal} {Phys. Rev. Lett.}\ }\textbf {\bibinfo {volume} {102}},\
  \bibinfo {pages} {247603} (\bibinfo {year} {2009})}\BibitemShut {NoStop}%
\bibitem [{\citenamefont {Subedi}(2015{\natexlab{a}})}]{subedi_2015}%
  \BibitemOpen
  \bibfield  {author} {\bibinfo {author} {\bibfnamefont {A.}~\bibnamefont
  {Subedi}},\ }\href {https://doi.org/10.1103/PhysRevB.92.214303} {\bibfield
  {journal} {\bibinfo  {journal} {Phys. Rev. B}\ }\textbf {\bibinfo {volume}
  {92}},\ \bibinfo {pages} {214303} (\bibinfo {year}
  {2015}{\natexlab{a}})}\BibitemShut {NoStop}%
\bibitem [{\citenamefont {Chen}\ \emph {et~al.}(2016)\citenamefont {Chen},
  \citenamefont {Zhu}, \citenamefont {Liu}, \citenamefont {Qi}, \citenamefont
  {Hwang}, \citenamefont {Brandt}, \citenamefont {Lu}, \citenamefont {Quirin},
  \citenamefont {Enquist}, \citenamefont {Zalden}, \citenamefont {Hu},
  \citenamefont {Goodfellow}, \citenamefont {Sher}, \citenamefont {Hoffmann},
  \citenamefont {Zhu}, \citenamefont {Lemke}, \citenamefont {Glownia},
  \citenamefont {Chollet}, \citenamefont {Damodaran}, \citenamefont {Park},
  \citenamefont {Cai}, \citenamefont {Jung}, \citenamefont {Highland},
  \citenamefont {Walko}, \citenamefont {Freeland}, \citenamefont {Evans},
  \citenamefont {Vailionis}, \citenamefont {Larsson}, \citenamefont {Nelson},
  \citenamefont {Rappe}, \citenamefont {Sokolowski-Tinten}, \citenamefont
  {Martin}, \citenamefont {Wen},\ and\ \citenamefont {Lindenberg}}]{chen_2016}%
  \BibitemOpen
  \bibfield  {author} {\bibinfo {author} {\bibfnamefont {F.}~\bibnamefont
  {Chen}}, \bibinfo {author} {\bibfnamefont {Y.}~\bibnamefont {Zhu}}, \bibinfo
  {author} {\bibfnamefont {S.}~\bibnamefont {Liu}}, \bibinfo {author}
  {\bibfnamefont {Y.}~\bibnamefont {Qi}}, \bibinfo {author} {\bibfnamefont
  {H.~Y.}\ \bibnamefont {Hwang}}, \bibinfo {author} {\bibfnamefont {N.~C.}\
  \bibnamefont {Brandt}}, \bibinfo {author} {\bibfnamefont {J.}~\bibnamefont
  {Lu}}, \bibinfo {author} {\bibfnamefont {F.}~\bibnamefont {Quirin}}, \bibinfo
  {author} {\bibfnamefont {H.}~\bibnamefont {Enquist}}, \bibinfo {author}
  {\bibfnamefont {P.}~\bibnamefont {Zalden}}, \bibinfo {author} {\bibfnamefont
  {T.}~\bibnamefont {Hu}}, \bibinfo {author} {\bibfnamefont {J.}~\bibnamefont
  {Goodfellow}}, \bibinfo {author} {\bibfnamefont {M.-J.}\ \bibnamefont
  {Sher}}, \bibinfo {author} {\bibfnamefont {M.~C.}\ \bibnamefont {Hoffmann}},
  \bibinfo {author} {\bibfnamefont {D.}~\bibnamefont {Zhu}}, \bibinfo {author}
  {\bibfnamefont {H.}~\bibnamefont {Lemke}}, \bibinfo {author} {\bibfnamefont
  {J.}~\bibnamefont {Glownia}}, \bibinfo {author} {\bibfnamefont
  {M.}~\bibnamefont {Chollet}}, \bibinfo {author} {\bibfnamefont {A.~R.}\
  \bibnamefont {Damodaran}}, \bibinfo {author} {\bibfnamefont {J.}~\bibnamefont
  {Park}}, \bibinfo {author} {\bibfnamefont {Z.}~\bibnamefont {Cai}}, \bibinfo
  {author} {\bibfnamefont {I.~W.}\ \bibnamefont {Jung}}, \bibinfo {author}
  {\bibfnamefont {M.~J.}\ \bibnamefont {Highland}}, \bibinfo {author}
  {\bibfnamefont {D.~A.}\ \bibnamefont {Walko}}, \bibinfo {author}
  {\bibfnamefont {J.~W.}\ \bibnamefont {Freeland}}, \bibinfo {author}
  {\bibfnamefont {P.~G.}\ \bibnamefont {Evans}}, \bibinfo {author}
  {\bibfnamefont {A.}~\bibnamefont {Vailionis}}, \bibinfo {author}
  {\bibfnamefont {J.}~\bibnamefont {Larsson}}, \bibinfo {author} {\bibfnamefont
  {K.~A.}\ \bibnamefont {Nelson}}, \bibinfo {author} {\bibfnamefont {A.~M.}\
  \bibnamefont {Rappe}}, \bibinfo {author} {\bibfnamefont {K.}~\bibnamefont
  {Sokolowski-Tinten}}, \bibinfo {author} {\bibfnamefont {L.~W.}\ \bibnamefont
  {Martin}}, \bibinfo {author} {\bibfnamefont {H.}~\bibnamefont {Wen}},\ and\
  \bibinfo {author} {\bibfnamefont {A.~M.}\ \bibnamefont {Lindenberg}},\ }\href
  {https://doi.org/10.1103/PhysRevB.94.180104} {\bibfield  {journal} {\bibinfo
  {journal} {Phys. Rev. B}\ }\textbf {\bibinfo {volume} {94}},\ \bibinfo
  {pages} {180104} (\bibinfo {year} {2016})}\BibitemShut {NoStop}%
\bibitem [{\citenamefont {Mankowsky}\ \emph {et~al.}(2017)\citenamefont
  {Mankowsky}, \citenamefont {von Hoegen}, \citenamefont {F\"orst},\ and\
  \citenamefont {Cavalleri}}]{mankowsky_2017}%
  \BibitemOpen
  \bibfield  {author} {\bibinfo {author} {\bibfnamefont {R.}~\bibnamefont
  {Mankowsky}}, \bibinfo {author} {\bibfnamefont {A.}~\bibnamefont {von
  Hoegen}}, \bibinfo {author} {\bibfnamefont {M.}~\bibnamefont {F\"orst}},\
  and\ \bibinfo {author} {\bibfnamefont {A.}~\bibnamefont {Cavalleri}},\ }\href
  {https://doi.org/10.1103/PhysRevLett.118.197601} {\bibfield  {journal}
  {\bibinfo  {journal} {Phys. Rev. Lett.}\ }\textbf {\bibinfo {volume} {118}},\
  \bibinfo {pages} {197601} (\bibinfo {year} {2017})}\BibitemShut {NoStop}%
\bibitem [{\citenamefont {Akamatsu}\ \emph {et~al.}(2018)\citenamefont
  {Akamatsu}, \citenamefont {Yuan}, \citenamefont {Stoica}, \citenamefont
  {Stone}, \citenamefont {Yang}, \citenamefont {Hong}, \citenamefont {Lei},
  \citenamefont {Zhu}, \citenamefont {Haislmaier}, \citenamefont {Freeland},
  \citenamefont {Chen}, \citenamefont {Wen},\ and\ \citenamefont
  {Gopalan}}]{freeland_2018}%
  \BibitemOpen
  \bibfield  {author} {\bibinfo {author} {\bibfnamefont {H.}~\bibnamefont
  {Akamatsu}}, \bibinfo {author} {\bibfnamefont {Y.}~\bibnamefont {Yuan}},
  \bibinfo {author} {\bibfnamefont {V.~A.}\ \bibnamefont {Stoica}}, \bibinfo
  {author} {\bibfnamefont {G.}~\bibnamefont {Stone}}, \bibinfo {author}
  {\bibfnamefont {T.}~\bibnamefont {Yang}}, \bibinfo {author} {\bibfnamefont
  {Z.}~\bibnamefont {Hong}}, \bibinfo {author} {\bibfnamefont {S.}~\bibnamefont
  {Lei}}, \bibinfo {author} {\bibfnamefont {Y.}~\bibnamefont {Zhu}}, \bibinfo
  {author} {\bibfnamefont {R.~C.}\ \bibnamefont {Haislmaier}}, \bibinfo
  {author} {\bibfnamefont {J.~W.}\ \bibnamefont {Freeland}}, \bibinfo {author}
  {\bibfnamefont {L.}~\bibnamefont {Chen}}, \bibinfo {author} {\bibfnamefont
  {H.}~\bibnamefont {Wen}},\ and\ \bibinfo {author} {\bibfnamefont
  {V.}~\bibnamefont {Gopalan}},\ }\href
  {https://doi.org/10.1103/PhysRevLett.120.096101} {\bibfield  {journal}
  {\bibinfo  {journal} {Phys. Rev. Lett.}\ }\textbf {\bibinfo {volume} {120}},\
  \bibinfo {pages} {096101} (\bibinfo {year} {2018})}\BibitemShut {NoStop}%
\bibitem [{\citenamefont {Rubio-Marcos}\ \emph {et~al.}(2018)\citenamefont
  {Rubio-Marcos}, \citenamefont {Ochoa}, \citenamefont {Del~Campo},
  \citenamefont {Garcia}, \citenamefont {Castro}, \citenamefont {Fernandez},\
  and\ \citenamefont {Garcia}}]{garcia_2018}%
  \BibitemOpen
  \bibfield  {author} {\bibinfo {author} {\bibfnamefont {F.}~\bibnamefont
  {Rubio-Marcos}}, \bibinfo {author} {\bibfnamefont {D.~A.}\ \bibnamefont
  {Ochoa}}, \bibinfo {author} {\bibfnamefont {A.}~\bibnamefont {Del~Campo}},
  \bibinfo {author} {\bibfnamefont {M.~A.}\ \bibnamefont {Garcia}}, \bibinfo
  {author} {\bibfnamefont {G.~R.}\ \bibnamefont {Castro}}, \bibinfo {author}
  {\bibfnamefont {J.~F.}\ \bibnamefont {Fernandez}},\ and\ \bibinfo {author}
  {\bibfnamefont {J.~E.}\ \bibnamefont {Garcia}},\ }\href
  {https://doi.org/10.1038/s41566-017-0068-1} {\bibfield  {journal} {\bibinfo
  {journal} {Nat. Phot.}\ }\textbf {\bibinfo {volume} {12}},\ \bibinfo {pages}
  {29} (\bibinfo {year} {2018})}\BibitemShut {NoStop}%
\bibitem [{\citenamefont {Li}\ \emph {et~al.}(2018)\citenamefont {Li},
  \citenamefont {Lipatov}, \citenamefont {Lu}, \citenamefont {Lee},
  \citenamefont {Lee}, \citenamefont {Torun}, \citenamefont {Wirtz},
  \citenamefont {Eom}, \citenamefont {Iniguez}, \citenamefont {Sinitskii},\
  and\ \citenamefont {Gruverman}}]{gruverman_2018}%
  \BibitemOpen
  \bibfield  {author} {\bibinfo {author} {\bibfnamefont {T.}~\bibnamefont
  {Li}}, \bibinfo {author} {\bibfnamefont {A.}~\bibnamefont {Lipatov}},
  \bibinfo {author} {\bibfnamefont {H.}~\bibnamefont {Lu}}, \bibinfo {author}
  {\bibfnamefont {H.}~\bibnamefont {Lee}}, \bibinfo {author} {\bibfnamefont
  {J.-W.}\ \bibnamefont {Lee}}, \bibinfo {author} {\bibfnamefont
  {E.}~\bibnamefont {Torun}}, \bibinfo {author} {\bibfnamefont
  {L.}~\bibnamefont {Wirtz}}, \bibinfo {author} {\bibfnamefont {C.-B.}\
  \bibnamefont {Eom}}, \bibinfo {author} {\bibfnamefont {J.}~\bibnamefont
  {Iniguez}}, \bibinfo {author} {\bibfnamefont {A.}~\bibnamefont {Sinitskii}},\
  and\ \bibinfo {author} {\bibfnamefont {A.}~\bibnamefont {Gruverman}},\
  }\href@noop {} {\bibfield  {journal} {\bibinfo  {journal} {Nat. Comm.}\
  }\textbf {\bibinfo {volume} {9}} (\bibinfo {year} {2018})}\BibitemShut
  {NoStop}%
\bibitem [{\citenamefont {Chen}\ \emph {et~al.}(2022)\citenamefont {Chen},
  \citenamefont {Paillard}, \citenamefont {Zhao}, \citenamefont {Iniguez},\
  and\ \citenamefont {Bellaiche}}]{Peng_2022}%
  \BibitemOpen
  \bibfield  {author} {\bibinfo {author} {\bibfnamefont {P.}~\bibnamefont
  {Chen}}, \bibinfo {author} {\bibfnamefont {C.}~\bibnamefont {Paillard}},
  \bibinfo {author} {\bibfnamefont {H.~J.}\ \bibnamefont {Zhao}}, \bibinfo
  {author} {\bibfnamefont {J.}~\bibnamefont {Iniguez}},\ and\ \bibinfo {author}
  {\bibfnamefont {L.}~\bibnamefont {Bellaiche}},\ }\href
  {https://doi.org/10.1038/s41467-022-30324-5} {\bibfield  {journal} {\bibinfo
  {journal} {Nat. Comm.}\ }\textbf {\bibinfo {volume} {13}},\ \bibinfo {pages}
  {2566} (\bibinfo {year} {2022})}\BibitemShut {NoStop}%
\bibitem [{\citenamefont {Gao}\ \emph {et~al.}(2023)\citenamefont {Gao},
  \citenamefont {Paillard},\ and\ \citenamefont
  {Bellaiche}}]{Bellaiche_PRB_2023}%
  \BibitemOpen
  \bibfield  {author} {\bibinfo {author} {\bibfnamefont {L.}~\bibnamefont
  {Gao}}, \bibinfo {author} {\bibfnamefont {C.}~\bibnamefont {Paillard}},\ and\
  \bibinfo {author} {\bibfnamefont {L.}~\bibnamefont {Bellaiche}},\ }\href
  {https://doi.org/10.1103/PhysRevB.107.104109} {\bibfield  {journal} {\bibinfo
   {journal} {Phys. Rev. B}\ }\textbf {\bibinfo {volume} {107}},\ \bibinfo
  {pages} {104109} (\bibinfo {year} {2023})}\BibitemShut {NoStop}%
\bibitem [{\citenamefont {Cavalleri}\ \emph {et~al.}(2006)\citenamefont
  {Cavalleri}, \citenamefont {Wall}, \citenamefont {Simpson}, \citenamefont
  {Statz}, \citenamefont {Ward}, \citenamefont {Nelson}, \citenamefont {Rini},\
  and\ \citenamefont {Schoenlein}}]{cavalleri_2006}%
  \BibitemOpen
  \bibfield  {author} {\bibinfo {author} {\bibfnamefont {A.}~\bibnamefont
  {Cavalleri}}, \bibinfo {author} {\bibfnamefont {S.}~\bibnamefont {Wall}},
  \bibinfo {author} {\bibfnamefont {C.}~\bibnamefont {Simpson}}, \bibinfo
  {author} {\bibfnamefont {E.}~\bibnamefont {Statz}}, \bibinfo {author}
  {\bibfnamefont {D.~W.}\ \bibnamefont {Ward}}, \bibinfo {author}
  {\bibfnamefont {K.~A.}\ \bibnamefont {Nelson}}, \bibinfo {author}
  {\bibfnamefont {M.}~\bibnamefont {Rini}},\ and\ \bibinfo {author}
  {\bibfnamefont {R.~W.}\ \bibnamefont {Schoenlein}},\ }\href
  {https://doi.org/10.1038/nature05041} {\bibfield  {journal} {\bibinfo
  {journal} {Nature}\ }\textbf {\bibinfo {volume} {442}},\ \bibinfo {pages}
  {664} (\bibinfo {year} {2006})}\BibitemShut {NoStop}%
\bibitem [{\citenamefont {Istomin}\ \emph {et~al.}(2007)\citenamefont
  {Istomin}, \citenamefont {Kotaidis}, \citenamefont {Plech},\ and\
  \citenamefont {Kong}}]{plech_2007}%
  \BibitemOpen
  \bibfield  {author} {\bibinfo {author} {\bibfnamefont {K.}~\bibnamefont
  {Istomin}}, \bibinfo {author} {\bibfnamefont {V.}~\bibnamefont {Kotaidis}},
  \bibinfo {author} {\bibfnamefont {A.}~\bibnamefont {Plech}},\ and\ \bibinfo
  {author} {\bibfnamefont {Q.}~\bibnamefont {Kong}},\ }\href
  {https://doi.org/10.1063/1.2430773} {\bibfield  {journal} {\bibinfo
  {journal} {Appl. Phys. Lett.}\ }\textbf {\bibinfo {volume} {90}},\ \bibinfo
  {pages} {022905} (\bibinfo {year} {2007})}\BibitemShut {NoStop}%
\bibitem [{\citenamefont {Rubio-Marcos}\ \emph {et~al.}(2015)\citenamefont
  {Rubio-Marcos}, \citenamefont {Del~Campo}, \citenamefont {Marchet},\ and\
  \citenamefont {Fern{\'a}ndez}}]{fernandez_2015}%
  \BibitemOpen
  \bibfield  {author} {\bibinfo {author} {\bibfnamefont {F.}~\bibnamefont
  {Rubio-Marcos}}, \bibinfo {author} {\bibfnamefont {A.}~\bibnamefont
  {Del~Campo}}, \bibinfo {author} {\bibfnamefont {P.}~\bibnamefont {Marchet}},\
  and\ \bibinfo {author} {\bibfnamefont {J.~F.}\ \bibnamefont
  {Fern{\'a}ndez}},\ }\href {https://doi.org/10.1038/ncomms7594} {\bibfield
  {journal} {\bibinfo  {journal} {Nat. Comm.}\ }\textbf {\bibinfo {volume}
  {6}},\ \bibinfo {pages} {6594} (\bibinfo {year} {2015})}\BibitemShut
  {NoStop}%
\bibitem [{\citenamefont {Bagri}\ \emph {et~al.}(2022)\citenamefont {Bagri},
  \citenamefont {Jana}, \citenamefont {Panchal}, \citenamefont {Phase},\ and\
  \citenamefont {Choudhary}}]{Bagri_2022}%
  \BibitemOpen
  \bibfield  {author} {\bibinfo {author} {\bibfnamefont {A.}~\bibnamefont
  {Bagri}}, \bibinfo {author} {\bibfnamefont {A.}~\bibnamefont {Jana}},
  \bibinfo {author} {\bibfnamefont {G.}~\bibnamefont {Panchal}}, \bibinfo
  {author} {\bibfnamefont {D.~M.}\ \bibnamefont {Phase}},\ and\ \bibinfo
  {author} {\bibfnamefont {R.~J.}\ \bibnamefont {Choudhary}},\ }\href
  {https://doi.org/10.1021/acsaelm.2c00694} {\bibfield  {journal} {\bibinfo
  {journal} {ACS Applied Electronic Materials}\ }\textbf {\bibinfo {volume}
  {4}},\ \bibinfo {pages} {4438} (\bibinfo {year} {2022})}\BibitemShut
  {NoStop}%
\bibitem [{\citenamefont {Tangney}\ and\ \citenamefont
  {Fahy}(1999)}]{Tangney_1999}%
  \BibitemOpen
  \bibfield  {author} {\bibinfo {author} {\bibfnamefont {P.}~\bibnamefont
  {Tangney}}\ and\ \bibinfo {author} {\bibfnamefont {S.}~\bibnamefont {Fahy}},\
  }\href {https://doi.org/10.1103/PhysRevLett.82.4340} {\bibfield  {journal}
  {\bibinfo  {journal} {Phys. Rev. Lett.}\ }\textbf {\bibinfo {volume} {82}},\
  \bibinfo {pages} {4340} (\bibinfo {year} {1999})}\BibitemShut {NoStop}%
\bibitem [{\citenamefont {Tangney}\ and\ \citenamefont
  {Fahy}(2002)}]{Tangney_2002}%
  \BibitemOpen
  \bibfield  {author} {\bibinfo {author} {\bibfnamefont {P.}~\bibnamefont
  {Tangney}}\ and\ \bibinfo {author} {\bibfnamefont {S.}~\bibnamefont {Fahy}},\
  }\href {https://doi.org/10.1103/PhysRevB.65.054302} {\bibfield  {journal}
  {\bibinfo  {journal} {Phys. Rev. B}\ }\textbf {\bibinfo {volume} {65}},\
  \bibinfo {pages} {054302} (\bibinfo {year} {2002})}\BibitemShut {NoStop}%
\bibitem [{\citenamefont {Murray}\ \emph {et~al.}(2005)\citenamefont {Murray},
  \citenamefont {Fritz}, \citenamefont {Wahlstrand}, \citenamefont {Fahy},\
  and\ \citenamefont {Reis}}]{Murray_2005PRB}%
  \BibitemOpen
  \bibfield  {author} {\bibinfo {author} {\bibfnamefont {E.~D.}\ \bibnamefont
  {Murray}}, \bibinfo {author} {\bibfnamefont {D.~M.}\ \bibnamefont {Fritz}},
  \bibinfo {author} {\bibfnamefont {J.~K.}\ \bibnamefont {Wahlstrand}},
  \bibinfo {author} {\bibfnamefont {S.}~\bibnamefont {Fahy}},\ and\ \bibinfo
  {author} {\bibfnamefont {D.~A.}\ \bibnamefont {Reis}},\ }\href
  {https://doi.org/10.1103/PhysRevB.72.060301} {\bibfield  {journal} {\bibinfo
  {journal} {Phys. Rev. B}\ }\textbf {\bibinfo {volume} {72}},\ \bibinfo
  {pages} {060301} (\bibinfo {year} {2005})}\BibitemShut {NoStop}%
\bibitem [{\citenamefont {Zijlstra}\ \emph {et~al.}(2006)\citenamefont
  {Zijlstra}, \citenamefont {Tatarinova},\ and\ \citenamefont
  {Garcia}}]{Garcia_PRB_2006}%
  \BibitemOpen
  \bibfield  {author} {\bibinfo {author} {\bibfnamefont {E.~S.}\ \bibnamefont
  {Zijlstra}}, \bibinfo {author} {\bibfnamefont {L.~L.}\ \bibnamefont
  {Tatarinova}},\ and\ \bibinfo {author} {\bibfnamefont {M.~E.}\ \bibnamefont
  {Garcia}},\ }\href {https://doi.org/10.1103/PhysRevB.74.220301} {\bibfield
  {journal} {\bibinfo  {journal} {Phys. Rev. B}\ }\textbf {\bibinfo {volume}
  {74}},\ \bibinfo {pages} {220301} (\bibinfo {year} {2006})}\BibitemShut
  {NoStop}%
\bibitem [{\citenamefont {Murray}\ \emph {et~al.}(2007)\citenamefont {Murray},
  \citenamefont {Fahy}, \citenamefont {Prendergast}, \citenamefont {Ogitsu},
  \citenamefont {Fritz},\ and\ \citenamefont {Reis}}]{Murray_PRB_2007}%
  \BibitemOpen
  \bibfield  {author} {\bibinfo {author} {\bibfnamefont {E.~D.}\ \bibnamefont
  {Murray}}, \bibinfo {author} {\bibfnamefont {S.}~\bibnamefont {Fahy}},
  \bibinfo {author} {\bibfnamefont {D.}~\bibnamefont {Prendergast}}, \bibinfo
  {author} {\bibfnamefont {T.}~\bibnamefont {Ogitsu}}, \bibinfo {author}
  {\bibfnamefont {D.~M.}\ \bibnamefont {Fritz}},\ and\ \bibinfo {author}
  {\bibfnamefont {D.~A.}\ \bibnamefont {Reis}},\ }\href
  {https://doi.org/10.1103/PhysRevB.75.184301} {\bibfield  {journal} {\bibinfo
  {journal} {Phys. Rev. B}\ }\textbf {\bibinfo {volume} {75}},\ \bibinfo
  {pages} {184301} (\bibinfo {year} {2007})}\BibitemShut {NoStop}%
\bibitem [{\citenamefont {Fritz}\ \emph {et~al.}(2007)\citenamefont {Fritz},
  \citenamefont {Reis}, \citenamefont {Adams}, \citenamefont {Akre},
  \citenamefont {Arthur}, \citenamefont {Blome}, \citenamefont {Bucksbaum},
  \citenamefont {Cavalieri}, \citenamefont {Engemann}, \citenamefont {Fahy},
  \citenamefont {Falcone}, \citenamefont {Fuoss}, \citenamefont {Gaffney},
  \citenamefont {George}, \citenamefont {Hajdu}, \citenamefont {Hertlein},
  \citenamefont {Hillyard}, \citenamefont {Hoegen}, \citenamefont {Kammler},
  \citenamefont {Kaspar}, \citenamefont {Kienberger}, \citenamefont {Krejcik},
  \citenamefont {Lee}, \citenamefont {Lindenberg}, \citenamefont {McFarland},
  \citenamefont {Meyer}, \citenamefont {Montagne}, \citenamefont {Murray},
  \citenamefont {Nelson}, \citenamefont {Nicoul}, \citenamefont {Pahl},
  \citenamefont {Rudati}, \citenamefont {Schlarb}, \citenamefont {Siddons},
  \citenamefont {Sokolowski-Tinten}, \citenamefont {Tschentscher},
  \citenamefont {von~der Linde},\ and\ \citenamefont
  {Hastings}}]{fahy_science_2007}%
  \BibitemOpen
  \bibfield  {author} {\bibinfo {author} {\bibfnamefont {D.~M.}\ \bibnamefont
  {Fritz}}, \bibinfo {author} {\bibfnamefont {D.~A.}\ \bibnamefont {Reis}},
  \bibinfo {author} {\bibfnamefont {B.}~\bibnamefont {Adams}}, \bibinfo
  {author} {\bibfnamefont {R.~A.}\ \bibnamefont {Akre}}, \bibinfo {author}
  {\bibfnamefont {J.}~\bibnamefont {Arthur}}, \bibinfo {author} {\bibfnamefont
  {C.}~\bibnamefont {Blome}}, \bibinfo {author} {\bibfnamefont {P.~H.}\
  \bibnamefont {Bucksbaum}}, \bibinfo {author} {\bibfnamefont {A.~L.}\
  \bibnamefont {Cavalieri}}, \bibinfo {author} {\bibfnamefont {S.}~\bibnamefont
  {Engemann}}, \bibinfo {author} {\bibfnamefont {S.}~\bibnamefont {Fahy}},
  \bibinfo {author} {\bibfnamefont {R.~W.}\ \bibnamefont {Falcone}}, \bibinfo
  {author} {\bibfnamefont {P.~H.}\ \bibnamefont {Fuoss}}, \bibinfo {author}
  {\bibfnamefont {K.~J.}\ \bibnamefont {Gaffney}}, \bibinfo {author}
  {\bibfnamefont {M.~J.}\ \bibnamefont {George}}, \bibinfo {author}
  {\bibfnamefont {J.}~\bibnamefont {Hajdu}}, \bibinfo {author} {\bibfnamefont
  {M.~P.}\ \bibnamefont {Hertlein}}, \bibinfo {author} {\bibfnamefont {P.~B.}\
  \bibnamefont {Hillyard}}, \bibinfo {author} {\bibfnamefont {M.~H.-v.}\
  \bibnamefont {Hoegen}}, \bibinfo {author} {\bibfnamefont {M.}~\bibnamefont
  {Kammler}}, \bibinfo {author} {\bibfnamefont {J.}~\bibnamefont {Kaspar}},
  \bibinfo {author} {\bibfnamefont {R.}~\bibnamefont {Kienberger}}, \bibinfo
  {author} {\bibfnamefont {P.}~\bibnamefont {Krejcik}}, \bibinfo {author}
  {\bibfnamefont {S.~H.}\ \bibnamefont {Lee}}, \bibinfo {author} {\bibfnamefont
  {A.~M.}\ \bibnamefont {Lindenberg}}, \bibinfo {author} {\bibfnamefont
  {B.}~\bibnamefont {McFarland}}, \bibinfo {author} {\bibfnamefont
  {D.}~\bibnamefont {Meyer}}, \bibinfo {author} {\bibfnamefont
  {T.}~\bibnamefont {Montagne}}, \bibinfo {author} {\bibfnamefont {E.~D.}\
  \bibnamefont {Murray}}, \bibinfo {author} {\bibfnamefont {A.~J.}\
  \bibnamefont {Nelson}}, \bibinfo {author} {\bibfnamefont {M.}~\bibnamefont
  {Nicoul}}, \bibinfo {author} {\bibfnamefont {R.}~\bibnamefont {Pahl}},
  \bibinfo {author} {\bibfnamefont {J.}~\bibnamefont {Rudati}}, \bibinfo
  {author} {\bibfnamefont {H.}~\bibnamefont {Schlarb}}, \bibinfo {author}
  {\bibfnamefont {D.~P.}\ \bibnamefont {Siddons}}, \bibinfo {author}
  {\bibfnamefont {K.}~\bibnamefont {Sokolowski-Tinten}}, \bibinfo {author}
  {\bibfnamefont {T.}~\bibnamefont {Tschentscher}}, \bibinfo {author}
  {\bibfnamefont {D.}~\bibnamefont {von~der Linde}},\ and\ \bibinfo {author}
  {\bibfnamefont {J.~B.}\ \bibnamefont {Hastings}},\ }\href
  {https://doi.org/10.1126/science.1135009} {\bibfield  {journal} {\bibinfo
  {journal} {Science}\ }\textbf {\bibinfo {volume} {315}},\ \bibinfo {pages}
  {633} (\bibinfo {year} {2007})}\BibitemShut {NoStop}%
\bibitem [{\citenamefont {Gu}\ \emph {et~al.}(2021)\citenamefont {Gu},
  \citenamefont {Murray},\ and\ \citenamefont {Tangney}}]{Gu2021}%
  \BibitemOpen
  \bibfield  {author} {\bibinfo {author} {\bibfnamefont {F.}~\bibnamefont
  {Gu}}, \bibinfo {author} {\bibfnamefont {E.}~\bibnamefont {Murray}},\ and\
  \bibinfo {author} {\bibfnamefont {P.}~\bibnamefont {Tangney}},\ }\href
  {https://doi.org/10.1103/PhysRevMaterials.5.034414} {\bibfield  {journal}
  {\bibinfo  {journal} {Phys. Rev. Mater.}\ }\textbf {\bibinfo {volume} {5}},\
  \bibinfo {pages} {034414} (\bibinfo {year} {2021})}\BibitemShut {NoStop}%
\bibitem [{\citenamefont {Sundaram}\ and\ \citenamefont
  {Mazur}(2002)}]{Sundaram2002}%
  \BibitemOpen
  \bibfield  {author} {\bibinfo {author} {\bibfnamefont {S.~K.}\ \bibnamefont
  {Sundaram}}\ and\ \bibinfo {author} {\bibfnamefont {E.}~\bibnamefont
  {Mazur}},\ }\href {https://doi.org/10.1038/nmat767} {\bibfield  {journal}
  {\bibinfo  {journal} {Nat. Mater.}\ }\textbf {\bibinfo {volume} {1}},\
  \bibinfo {pages} {217} (\bibinfo {year} {2002})}\BibitemShut {NoStop}%
\bibitem [{\citenamefont {Gamaly}(2011)}]{Gamaly_2011}%
  \BibitemOpen
  \bibfield  {author} {\bibinfo {author} {\bibfnamefont {E.}~\bibnamefont
  {Gamaly}},\ }\href
  {https://doi.org/https://doi.org/10.1016/j.physrep.2011.07.002} {\bibfield
  {journal} {\bibinfo  {journal} {Phys. Rep.}\ }\textbf {\bibinfo {volume}
  {508}},\ \bibinfo {pages} {91} (\bibinfo {year} {2011})}\BibitemShut
  {NoStop}%
\bibitem [{\citenamefont {Shah}(2013)}]{shah_2013}%
  \BibitemOpen
  \bibfield  {author} {\bibinfo {author} {\bibfnamefont {J.}~\bibnamefont
  {Shah}},\ }\href {https://books.google.is/books?id=Q_H1CAAAQBAJ} {\emph
  {\bibinfo {title} {Ultrafast Spectroscopy of Semiconductors and Semiconductor
  Nanostructures}}},\ Springer Series in Solid-State Sciences\ (\bibinfo
  {publisher} {Springer Berlin Heidelberg},\ \bibinfo {year}
  {2013})\BibitemShut {NoStop}%
\bibitem [{\citenamefont {Phillips}\ \emph {et~al.}(2015)\citenamefont
  {Phillips}, \citenamefont {Gandhi}, \citenamefont {Mazur},\ and\
  \citenamefont {Sundaram}}]{Phillips_2015}%
  \BibitemOpen
  \bibfield  {author} {\bibinfo {author} {\bibfnamefont {K.~C.}\ \bibnamefont
  {Phillips}}, \bibinfo {author} {\bibfnamefont {H.~H.}\ \bibnamefont
  {Gandhi}}, \bibinfo {author} {\bibfnamefont {E.}~\bibnamefont {Mazur}},\ and\
  \bibinfo {author} {\bibfnamefont {S.~K.}\ \bibnamefont {Sundaram}},\ }\href
  {https://doi.org/10.1364/AOP.7.000684} {\bibfield  {journal} {\bibinfo
  {journal} {Adv. Opt. Photon.}\ }\textbf {\bibinfo {volume} {7}},\ \bibinfo
  {pages} {684} (\bibinfo {year} {2015})}\BibitemShut {NoStop}%
\bibitem [{\citenamefont {Zeiger}\ \emph {et~al.}(1992)\citenamefont {Zeiger},
  \citenamefont {Vidal}, \citenamefont {Cheng}, \citenamefont {Ippen},
  \citenamefont {Dresselhaus},\ and\ \citenamefont {Dresselhaus}}]{zeiger1992}%
  \BibitemOpen
  \bibfield  {author} {\bibinfo {author} {\bibfnamefont {H.~J.}\ \bibnamefont
  {Zeiger}}, \bibinfo {author} {\bibfnamefont {J.}~\bibnamefont {Vidal}},
  \bibinfo {author} {\bibfnamefont {T.~K.}\ \bibnamefont {Cheng}}, \bibinfo
  {author} {\bibfnamefont {E.~P.}\ \bibnamefont {Ippen}}, \bibinfo {author}
  {\bibfnamefont {G.}~\bibnamefont {Dresselhaus}},\ and\ \bibinfo {author}
  {\bibfnamefont {M.~S.}\ \bibnamefont {Dresselhaus}},\ }\href
  {https://doi.org/10.1103/PhysRevB.45.768} {\bibfield  {journal} {\bibinfo
  {journal} {Phys. Rev. B}\ }\textbf {\bibinfo {volume} {45}},\ \bibinfo
  {pages} {768} (\bibinfo {year} {1992})}\BibitemShut {NoStop}%
\bibitem [{\citenamefont {Hunsche}\ \emph {et~al.}(1995)\citenamefont
  {Hunsche}, \citenamefont {Wienecke}, \citenamefont {Dekorsy},\ and\
  \citenamefont {Kurz}}]{Hunsche_PRL1995}%
  \BibitemOpen
  \bibfield  {author} {\bibinfo {author} {\bibfnamefont {S.}~\bibnamefont
  {Hunsche}}, \bibinfo {author} {\bibfnamefont {K.}~\bibnamefont {Wienecke}},
  \bibinfo {author} {\bibfnamefont {T.}~\bibnamefont {Dekorsy}},\ and\ \bibinfo
  {author} {\bibfnamefont {H.}~\bibnamefont {Kurz}},\ }\href
  {https://doi.org/10.1103/PhysRevLett.75.1815} {\bibfield  {journal} {\bibinfo
   {journal} {Phys. Rev. Lett.}\ }\textbf {\bibinfo {volume} {75}},\ \bibinfo
  {pages} {1815} (\bibinfo {year} {1995})}\BibitemShut {NoStop}%
\bibitem [{\citenamefont {Bargheer}\ \emph {et~al.}(2004)\citenamefont
  {Bargheer}, \citenamefont {Zhavoronkov}, \citenamefont {Gritsai},
  \citenamefont {Woo}, \citenamefont {Kim}, \citenamefont {W{\"o}rner},\ and\
  \citenamefont {Els{\"a}sser}}]{bargheer2004}%
  \BibitemOpen
  \bibfield  {author} {\bibinfo {author} {\bibfnamefont {M.}~\bibnamefont
  {Bargheer}}, \bibinfo {author} {\bibfnamefont {N.}~\bibnamefont
  {Zhavoronkov}}, \bibinfo {author} {\bibfnamefont {Y.}~\bibnamefont
  {Gritsai}}, \bibinfo {author} {\bibfnamefont {J.}~\bibnamefont {Woo}},
  \bibinfo {author} {\bibfnamefont {D.}~\bibnamefont {Kim}}, \bibinfo {author}
  {\bibfnamefont {M.}~\bibnamefont {W{\"o}rner}},\ and\ \bibinfo {author}
  {\bibfnamefont {T.}~\bibnamefont {Els{\"a}sser}},\ }\href@noop {} {\bibfield
  {journal} {\bibinfo  {journal} {Science}\ }\textbf {\bibinfo {volume}
  {306}},\ \bibinfo {pages} {1771} (\bibinfo {year} {2004})}\BibitemShut
  {NoStop}%
\bibitem [{\citenamefont {Axe}(1967)}]{axe_1967}%
  \BibitemOpen
  \bibfield  {author} {\bibinfo {author} {\bibfnamefont {J.~D.}\ \bibnamefont
  {Axe}},\ }\href {https://doi.org/10.1103/PhysRev.157.429} {\bibfield
  {journal} {\bibinfo  {journal} {Phys. Rev.}\ }\textbf {\bibinfo {volume}
  {157}},\ \bibinfo {pages} {429} (\bibinfo {year} {1967})}\BibitemShut
  {NoStop}%
\bibitem [{\citenamefont {Last}(1957)}]{last_1957}%
  \BibitemOpen
  \bibfield  {author} {\bibinfo {author} {\bibfnamefont {J.~T.}\ \bibnamefont
  {Last}},\ }\href {https://doi.org/10.1103/PhysRev.105.1740} {\bibfield
  {journal} {\bibinfo  {journal} {Phys. Rev.}\ }\textbf {\bibinfo {volume}
  {105}},\ \bibinfo {pages} {1740} (\bibinfo {year} {1957})}\BibitemShut
  {NoStop}%
\bibitem [{\citenamefont {Slater}(1950)}]{slater_1950}%
  \BibitemOpen
  \bibfield  {author} {\bibinfo {author} {\bibfnamefont {J.~C.}\ \bibnamefont
  {Slater}},\ }\href {https://doi.org/10.1103/PhysRev.78.748} {\bibfield
  {journal} {\bibinfo  {journal} {Phys. Rev.}\ }\textbf {\bibinfo {volume}
  {78}},\ \bibinfo {pages} {748} (\bibinfo {year} {1950})}\BibitemShut
  {NoStop}%
\bibitem [{\citenamefont {Nemytov}(2019)}]{nemytov_thesis}%
  \BibitemOpen
  \bibfield  {author} {\bibinfo {author} {\bibfnamefont {V.}~\bibnamefont
  {Nemytov}},\ }\emph {\bibinfo {title} {Towards an accurate and transferable
  charge transfer model in polarisable interatomic potentials}},\ \href@noop {}
  {Ph.D. thesis},\ \bibinfo  {school} {Imperial College London} (\bibinfo
  {year} {2019})\BibitemShut {NoStop}%
\bibitem [{\citenamefont {Goldschmidt}(1926)}]{goldschmidt_ratio}%
  \BibitemOpen
  \bibfield  {author} {\bibinfo {author} {\bibfnamefont {V.~M.}\ \bibnamefont
  {Goldschmidt}},\ }\href {https://doi.org/10.1007/BF0150752} {\bibfield
  {journal} {\bibinfo  {journal} {Naturewissenschaften}\ }\textbf {\bibinfo
  {volume} {14}},\ \bibinfo {pages} {477} (\bibinfo {year} {1926})}\BibitemShut
  {NoStop}%
\bibitem [{\citenamefont {Zhong}\ \emph {et~al.}(1994)\citenamefont {Zhong},
  \citenamefont {Vanderbilt},\ and\ \citenamefont {Rabe}}]{zhong_PRL_1994}%
  \BibitemOpen
  \bibfield  {author} {\bibinfo {author} {\bibfnamefont {W.}~\bibnamefont
  {Zhong}}, \bibinfo {author} {\bibfnamefont {D.}~\bibnamefont {Vanderbilt}},\
  and\ \bibinfo {author} {\bibfnamefont {K.~M.}\ \bibnamefont {Rabe}},\ }\href
  {https://doi.org/10.1103/PhysRevLett.73.1861} {\bibfield  {journal} {\bibinfo
   {journal} {Phys. Rev. Lett.}\ }\textbf {\bibinfo {volume} {73}},\ \bibinfo
  {pages} {1861} (\bibinfo {year} {1994})}\BibitemShut {NoStop}%
\bibitem [{\citenamefont {Fallon}(2014)}]{fallon_thesis}%
  \BibitemOpen
  \bibfield  {author} {\bibinfo {author} {\bibfnamefont {J.}~\bibnamefont
  {Fallon}},\ }\emph {\bibinfo {title} {Multiscale theory and simulation of
  barium titanate}},\ \href@noop {} {Ph.D. thesis},\ \bibinfo  {school}
  {Imperial College London} (\bibinfo {year} {2014})\BibitemShut {NoStop}%
\bibitem [{\citenamefont {Hlinka}\ \emph {et~al.}(2008)\citenamefont {Hlinka},
  \citenamefont {Ostapchuk}, \citenamefont {Nuzhnyy}, \citenamefont {Petzelt},
  \citenamefont {Kuzel}, \citenamefont {Kadlec}, \citenamefont {Vanek},
  \citenamefont {Ponomareva},\ and\ \citenamefont
  {Bellaiche}}]{bellaiche_2008}%
  \BibitemOpen
  \bibfield  {author} {\bibinfo {author} {\bibfnamefont {J.}~\bibnamefont
  {Hlinka}}, \bibinfo {author} {\bibfnamefont {T.}~\bibnamefont {Ostapchuk}},
  \bibinfo {author} {\bibfnamefont {D.}~\bibnamefont {Nuzhnyy}}, \bibinfo
  {author} {\bibfnamefont {J.}~\bibnamefont {Petzelt}}, \bibinfo {author}
  {\bibfnamefont {P.}~\bibnamefont {Kuzel}}, \bibinfo {author} {\bibfnamefont
  {C.}~\bibnamefont {Kadlec}}, \bibinfo {author} {\bibfnamefont
  {P.}~\bibnamefont {Vanek}}, \bibinfo {author} {\bibfnamefont
  {I.}~\bibnamefont {Ponomareva}},\ and\ \bibinfo {author} {\bibfnamefont
  {L.}~\bibnamefont {Bellaiche}},\ }\href
  {https://doi.org/10.1103/PhysRevLett.101.167402} {\bibfield  {journal}
  {\bibinfo  {journal} {Phys. Rev. Lett.}\ }\textbf {\bibinfo {volume} {101}},\
  \bibinfo {pages} {167402} (\bibinfo {year} {2008})}\BibitemShut {NoStop}%
\bibitem [{\citenamefont {Weerasinghe}\ \emph {et~al.}(2013)\citenamefont
  {Weerasinghe}, \citenamefont {Bellaiche}, \citenamefont {Ostapchuk},
  \citenamefont {Kužel}, \citenamefont {Kadlec}, \citenamefont {Lisenkov},
  \citenamefont {Ponomareva},\ and\ \citenamefont
  {Hlinka}}]{weerasinghe_bellaiche_2013}%
  \BibitemOpen
  \bibfield  {author} {\bibinfo {author} {\bibfnamefont {J.}~\bibnamefont
  {Weerasinghe}}, \bibinfo {author} {\bibfnamefont {L.}~\bibnamefont
  {Bellaiche}}, \bibinfo {author} {\bibfnamefont {T.}~\bibnamefont
  {Ostapchuk}}, \bibinfo {author} {\bibfnamefont {P.}~\bibnamefont {Kužel}},
  \bibinfo {author} {\bibfnamefont {C.}~\bibnamefont {Kadlec}}, \bibinfo
  {author} {\bibfnamefont {S.}~\bibnamefont {Lisenkov}}, \bibinfo {author}
  {\bibfnamefont {I.}~\bibnamefont {Ponomareva}},\ and\ \bibinfo {author}
  {\bibfnamefont {J.}~\bibnamefont {Hlinka}},\ }\href
  {https://doi.org/10.1557/mrc.2013.5} {\bibfield  {journal} {\bibinfo
  {journal} {MRS Commun.}\ }\textbf {\bibinfo {volume} {3}},\ \bibinfo {pages}
  {41–45} (\bibinfo {year} {2013})}\BibitemShut {NoStop}%
\bibitem [{\citenamefont {Pirc}\ and\ \citenamefont
  {Blinc}(2004)}]{blinc_2004}%
  \BibitemOpen
  \bibfield  {author} {\bibinfo {author} {\bibfnamefont {R.}~\bibnamefont
  {Pirc}}\ and\ \bibinfo {author} {\bibfnamefont {R.}~\bibnamefont {Blinc}},\
  }\href {https://doi.org/10.1103/PhysRevB.70.134107} {\bibfield  {journal}
  {\bibinfo  {journal} {Phys. Rev. B}\ }\textbf {\bibinfo {volume} {70}},\
  \bibinfo {pages} {134107} (\bibinfo {year} {2004})}\BibitemShut {NoStop}%
\bibitem [{\citenamefont {Senn}\ \emph {et~al.}(2016)\citenamefont {Senn},
  \citenamefont {Keen}, \citenamefont {Lucas}, \citenamefont {Hriljac},\ and\
  \citenamefont {Goodwin}}]{Senn_PRL_2016}%
  \BibitemOpen
  \bibfield  {author} {\bibinfo {author} {\bibfnamefont {M.~S.}\ \bibnamefont
  {Senn}}, \bibinfo {author} {\bibfnamefont {D.~A.}\ \bibnamefont {Keen}},
  \bibinfo {author} {\bibfnamefont {T.~C.~A.}\ \bibnamefont {Lucas}}, \bibinfo
  {author} {\bibfnamefont {J.~A.}\ \bibnamefont {Hriljac}},\ and\ \bibinfo
  {author} {\bibfnamefont {A.~L.}\ \bibnamefont {Goodwin}},\ }\href
  {https://doi.org/10.1103/PhysRevLett.116.207602} {\bibfield  {journal}
  {\bibinfo  {journal} {Phys. Rev. Lett.}\ }\textbf {\bibinfo {volume} {116}},\
  \bibinfo {pages} {207602} (\bibinfo {year} {2016})}\BibitemShut {NoStop}%
\bibitem [{\citenamefont {Gigli}\ \emph {et~al.}(2022)\citenamefont {Gigli},
  \citenamefont {Veit}, \citenamefont {Kotiuga}, \citenamefont {Pizzi},
  \citenamefont {Marzari},\ and\ \citenamefont {Ceriotti}}]{ceriotti_2022}%
  \BibitemOpen
  \bibfield  {author} {\bibinfo {author} {\bibfnamefont {L.}~\bibnamefont
  {Gigli}}, \bibinfo {author} {\bibfnamefont {M.}~\bibnamefont {Veit}},
  \bibinfo {author} {\bibfnamefont {M.}~\bibnamefont {Kotiuga}}, \bibinfo
  {author} {\bibfnamefont {G.}~\bibnamefont {Pizzi}}, \bibinfo {author}
  {\bibfnamefont {N.}~\bibnamefont {Marzari}},\ and\ \bibinfo {author}
  {\bibfnamefont {M.}~\bibnamefont {Ceriotti}},\ }\href
  {https://doi.org/10.1038/s41524-022-00845-0} {\bibfield  {journal} {\bibinfo
  {journal} {npj {Comput.} {Mater,}}\ }\textbf {\bibinfo {volume} {8}},\
  \bibinfo {pages} {209} (\bibinfo {year} {2022})}\BibitemShut {NoStop}%
\bibitem [{\citenamefont {Bersuker}(1966)}]{bersuker_1966}%
  \BibitemOpen
  \bibfield  {author} {\bibinfo {author} {\bibfnamefont {I.~B.}\ \bibnamefont
  {Bersuker}},\ }\href
  {https://doi.org/https://doi.org/10.1016/0031-9163(66)91127-9} {\bibfield
  {journal} {\bibinfo  {journal} {Phys. Lett.}\ }\textbf {\bibinfo {volume}
  {20}},\ \bibinfo {pages} {589} (\bibinfo {year} {1966})}\BibitemShut
  {NoStop}%
\bibitem [{\citenamefont {Comes}\ \emph {et~al.}(1968)\citenamefont {Comes},
  \citenamefont {Lambert},\ and\ \citenamefont {Guinier}}]{comes_1968}%
  \BibitemOpen
  \bibfield  {author} {\bibinfo {author} {\bibfnamefont {R.}~\bibnamefont
  {Comes}}, \bibinfo {author} {\bibfnamefont {M.}~\bibnamefont {Lambert}},\
  and\ \bibinfo {author} {\bibfnamefont {A.}~\bibnamefont {Guinier}},\ }\href
  {https://doi.org/https://doi.org/10.1016/0038-1098(68)90571-1} {\bibfield
  {journal} {\bibinfo  {journal} {Solid State Commun.}\ }\textbf {\bibinfo
  {volume} {6}},\ \bibinfo {pages} {715} (\bibinfo {year} {1968})}\BibitemShut
  {NoStop}%
\bibitem [{\citenamefont {Chaves}\ \emph {et~al.}(1976)\citenamefont {Chaves},
  \citenamefont {Barreto}, \citenamefont {Nogueira},\ and\ \citenamefont
  {Z\~eks}}]{chaves_1976}%
  \BibitemOpen
  \bibfield  {author} {\bibinfo {author} {\bibfnamefont {A.~S.}\ \bibnamefont
  {Chaves}}, \bibinfo {author} {\bibfnamefont {F.~C.~S.}\ \bibnamefont
  {Barreto}}, \bibinfo {author} {\bibfnamefont {R.~A.}\ \bibnamefont
  {Nogueira}},\ and\ \bibinfo {author} {\bibfnamefont {B.}~\bibnamefont
  {Z\~eks}},\ }\href {https://doi.org/10.1103/PhysRevB.13.207} {\bibfield
  {journal} {\bibinfo  {journal} {Phys. Rev. B}\ }\textbf {\bibinfo {volume}
  {13}},\ \bibinfo {pages} {207} (\bibinfo {year} {1976})}\BibitemShut
  {NoStop}%
\bibitem [{\citenamefont {Subedi}(2015{\natexlab{b}})}]{Subedi_PRB_2015}%
  \BibitemOpen
  \bibfield  {author} {\bibinfo {author} {\bibfnamefont {A.}~\bibnamefont
  {Subedi}},\ }\href {https://doi.org/10.1103/PhysRevB.92.214303} {\bibfield
  {journal} {\bibinfo  {journal} {Phys. Rev. B}\ }\textbf {\bibinfo {volume}
  {92}},\ \bibinfo {pages} {214303} (\bibinfo {year}
  {2015}{\natexlab{b}})}\BibitemShut {NoStop}%
\bibitem [{\citenamefont {Gayathri}\ \emph {et~al.}(2023)\citenamefont
  {Gayathri}, \citenamefont {Swamynathan}, \citenamefont {Shaikh},
  \citenamefont {Ghosh},\ and\ \citenamefont {Ghosh}}]{Gayathri_2023}%
  \BibitemOpen
  \bibfield  {author} {\bibinfo {author} {\bibfnamefont {P.}~\bibnamefont
  {Gayathri}}, \bibinfo {author} {\bibfnamefont {M.~J.}\ \bibnamefont
  {Swamynathan}}, \bibinfo {author} {\bibfnamefont {M.}~\bibnamefont {Shaikh}},
  \bibinfo {author} {\bibfnamefont {A.}~\bibnamefont {Ghosh}},\ and\ \bibinfo
  {author} {\bibfnamefont {S.}~\bibnamefont {Ghosh}},\ }\href
  {https://doi.org/10.1021/acs.chemmater.3c00108} {\bibfield  {journal}
  {\bibinfo  {journal} {Chemistry of Materials}\ }\textbf {\bibinfo {volume}
  {35}},\ \bibinfo {pages} {6612} (\bibinfo {year} {2023})}\BibitemShut
  {NoStop}%
\bibitem [{\citenamefont {Tangney}\ and\ \citenamefont
  {Scandolo}(2003)}]{tangney_jcp_2003}%
  \BibitemOpen
  \bibfield  {author} {\bibinfo {author} {\bibfnamefont {P.}~\bibnamefont
  {Tangney}}\ and\ \bibinfo {author} {\bibfnamefont {S.}~\bibnamefont
  {Scandolo}},\ }\href {https://doi.org/10.1063/1.1609980} {\bibfield
  {journal} {\bibinfo  {journal} {J. Chem. Phys.}\ }\textbf {\bibinfo {volume}
  {119}},\ \bibinfo {pages} {9673} (\bibinfo {year} {2003})}\BibitemShut
  {NoStop}%
\bibitem [{\citenamefont {Sarsam}\ \emph {et~al.}(2013)\citenamefont {Sarsam},
  \citenamefont {Finnis},\ and\ \citenamefont {Tangney}}]{sarsam_jcp_2013}%
  \BibitemOpen
  \bibfield  {author} {\bibinfo {author} {\bibfnamefont {J.}~\bibnamefont
  {Sarsam}}, \bibinfo {author} {\bibfnamefont {M.~W.}\ \bibnamefont {Finnis}},\
  and\ \bibinfo {author} {\bibfnamefont {P.}~\bibnamefont {Tangney}},\ }\href
  {https://doi.org/10.1063/1.4832695} {\bibfield  {journal} {\bibinfo
  {journal} {J. Chem. Phys.}\ }\textbf {\bibinfo {volume} {139}},\ \bibinfo
  {pages} {204704} (\bibinfo {year} {2013})}\BibitemShut {NoStop}%
\bibitem [{\citenamefont {Morse}(1929)}]{Morse_1929}%
  \BibitemOpen
  \bibfield  {author} {\bibinfo {author} {\bibfnamefont {P.~M.}\ \bibnamefont
  {Morse}},\ }\href {https://doi.org/10.1103/PhysRev.34.57} {\bibfield
  {journal} {\bibinfo  {journal} {Phys. Rev.}\ }\textbf {\bibinfo {volume}
  {34}},\ \bibinfo {pages} {57} (\bibinfo {year} {1929})}\BibitemShut {NoStop}%
\bibitem [{\citenamefont {Demiralp}\ \emph {et~al.}(1999)\citenamefont
  {Demiralp}, \citenamefont {\ifmmode \mbox{\c{C}}\else
  \c{C}\fi{}a\ifmmode~\breve{g}\else \u{g}\fi{}in},\ and\ \citenamefont
  {Goddard}}]{Demiralp_1999}%
  \BibitemOpen
  \bibfield  {author} {\bibinfo {author} {\bibfnamefont {E.}~\bibnamefont
  {Demiralp}}, \bibinfo {author} {\bibfnamefont {T.}~\bibnamefont {\ifmmode
  \mbox{\c{C}}\else \c{C}\fi{}a\ifmmode~\breve{g}\else \u{g}\fi{}in}},\ and\
  \bibinfo {author} {\bibfnamefont {W.~A.}\ \bibnamefont {Goddard}},\ }\href
  {https://doi.org/10.1103/PhysRevLett.82.1708} {\bibfield  {journal} {\bibinfo
   {journal} {Phys. Rev. Lett.}\ }\textbf {\bibinfo {volume} {82}},\ \bibinfo
  {pages} {1708} (\bibinfo {year} {1999})}\BibitemShut {NoStop}%
\bibitem [{\citenamefont {Sarsam}(2013)}]{Sarsam2013}%
  \BibitemOpen
  \bibfield  {author} {\bibinfo {author} {\bibfnamefont {J.}~\bibnamefont
  {Sarsam}},\ }\emph {\bibinfo {title} {Development and application of
  atomistic force fields for ionic materials}},\ \href@noop {} {Ph.D. thesis},\
  \bibinfo  {school} {Imperial College London} (\bibinfo {year}
  {2013})\BibitemShut {NoStop}%
\bibitem [{\citenamefont {Perdew}\ \emph {et~al.}(2008)\citenamefont {Perdew},
  \citenamefont {Ruzsinszky}, \citenamefont {Csonka}, \citenamefont {Vydrov},
  \citenamefont {Scuseria}, \citenamefont {Constantin}, \citenamefont {Zhou},\
  and\ \citenamefont {Burke}}]{PBE-sol}%
  \BibitemOpen
  \bibfield  {author} {\bibinfo {author} {\bibfnamefont {J.~P.}\ \bibnamefont
  {Perdew}}, \bibinfo {author} {\bibfnamefont {A.}~\bibnamefont {Ruzsinszky}},
  \bibinfo {author} {\bibfnamefont {G.~I.}\ \bibnamefont {Csonka}}, \bibinfo
  {author} {\bibfnamefont {O.~A.}\ \bibnamefont {Vydrov}}, \bibinfo {author}
  {\bibfnamefont {G.~E.}\ \bibnamefont {Scuseria}}, \bibinfo {author}
  {\bibfnamefont {L.~A.}\ \bibnamefont {Constantin}}, \bibinfo {author}
  {\bibfnamefont {X.}~\bibnamefont {Zhou}},\ and\ \bibinfo {author}
  {\bibfnamefont {K.}~\bibnamefont {Burke}},\ }\href
  {https://doi.org/10.1103/PhysRevLett.100.136406} {\bibfield  {journal}
  {\bibinfo  {journal} {Phys. Rev. Lett.}\ }\textbf {\bibinfo {volume} {100}},\
  \bibinfo {pages} {136406} (\bibinfo {year} {2008})}\BibitemShut {NoStop}%
\bibitem [{\citenamefont {Kwei}\ \emph {et~al.}(1993)\citenamefont {Kwei},
  \citenamefont {Lawson}, \citenamefont {Billinge},\ and\ \citenamefont
  {Cheong}}]{kwei_JPhysChem_1993}%
  \BibitemOpen
  \bibfield  {author} {\bibinfo {author} {\bibfnamefont {G.~H.}\ \bibnamefont
  {Kwei}}, \bibinfo {author} {\bibfnamefont {A.~C.}\ \bibnamefont {Lawson}},
  \bibinfo {author} {\bibfnamefont {S.~J.~L.}\ \bibnamefont {Billinge}},\ and\
  \bibinfo {author} {\bibfnamefont {S.~W.}\ \bibnamefont {Cheong}},\ }\href
  {https://doi.org/10.1021/j100112a043} {\bibfield  {journal} {\bibinfo
  {journal} {J. Phys. Chem.}\ }\textbf {\bibinfo {volume} {97}},\ \bibinfo
  {pages} {2368} (\bibinfo {year} {1993})}\BibitemShut {NoStop}%
\bibitem [{\citenamefont {Hellwege}\ \emph {et~al.}(1982)\citenamefont
  {Hellwege}, \citenamefont {Mitsui},\ and\ \citenamefont
  {Hellwege}}]{hellwege1982ferro}%
  \BibitemOpen
  \bibfield  {author} {\bibinfo {author} {\bibfnamefont {K.-H.}\ \bibnamefont
  {Hellwege}}, \bibinfo {author} {\bibfnamefont {T.}~\bibnamefont {Mitsui}},\
  and\ \bibinfo {author} {\bibfnamefont {A.~M.}\ \bibnamefont {Hellwege}},\
  }\href@noop {} {\emph {\bibinfo {title} {Ferroelectrics and Related
  Substances.}}}\ (\bibinfo  {publisher} {Springer-Verlag},\ \bibinfo {year}
  {1982})\BibitemShut {NoStop}%
\bibitem [{\citenamefont {Togo}\ and\ \citenamefont {Tanaka}(2015)}]{phonopy}%
  \BibitemOpen
  \bibfield  {author} {\bibinfo {author} {\bibfnamefont {A.}~\bibnamefont
  {Togo}}\ and\ \bibinfo {author} {\bibfnamefont {I.}~\bibnamefont {Tanaka}},\
  }\href@noop {} {\bibfield  {journal} {\bibinfo  {journal} {Scr. Mater.}\
  }\textbf {\bibinfo {volume} {108}},\ \bibinfo {pages} {1} (\bibinfo {year}
  {2015})}\BibitemShut {NoStop}%
\bibitem [{\citenamefont {Murnaghan}(1944)}]{Murnaghan244}%
  \BibitemOpen
  \bibfield  {author} {\bibinfo {author} {\bibfnamefont {F.~D.}\ \bibnamefont
  {Murnaghan}},\ }\href {https://doi.org/10.1073/pnas.30.9.244} {\bibfield
  {journal} {\bibinfo  {journal} {Proceedings of the National Academy of
  Sciences}\ }\textbf {\bibinfo {volume} {30}},\ \bibinfo {pages} {244}
  (\bibinfo {year} {1944})}\BibitemShut {NoStop}%
\bibitem [{\citenamefont {Piskunov}(2004)}]{Piskunov2004}%
  \BibitemOpen
  \bibfield  {author} {\bibinfo {author} {\bibfnamefont {S.}~\bibnamefont
  {Piskunov}},\ }\href {https://doi.org/10.1016/j.commatsci.2003.08.036} {\
  \textbf {\bibinfo {volume} {29}},\ \bibinfo {pages} {165} (\bibinfo {year}
  {2004})}\BibitemShut {NoStop}%
\bibitem [{\citenamefont {King-Smith}\ and\ \citenamefont
  {Vanderbilt}(1994)}]{kingsmith1994}%
  \BibitemOpen
  \bibfield  {author} {\bibinfo {author} {\bibfnamefont {R.~D.}\ \bibnamefont
  {King-Smith}}\ and\ \bibinfo {author} {\bibfnamefont {D.}~\bibnamefont
  {Vanderbilt}},\ }\href {https://doi.org/10.1103/PhysRevB.49.5828} {\bibfield
  {journal} {\bibinfo  {journal} {Phys. Rev. B}\ }\textbf {\bibinfo {volume}
  {49}},\ \bibinfo {pages} {5828} (\bibinfo {year} {1994})}\BibitemShut
  {NoStop}%
\bibitem [{\citenamefont {Tinte}\ \emph {et~al.}(1999)\citenamefont {Tinte},
  \citenamefont {Stachiotti}, \citenamefont {Sepliarsky}, \citenamefont
  {Migoni},\ and\ \citenamefont {Rodriguez}}]{Tinte_1999}%
  \BibitemOpen
  \bibfield  {author} {\bibinfo {author} {\bibfnamefont {S.}~\bibnamefont
  {Tinte}}, \bibinfo {author} {\bibfnamefont {M.~G.}\ \bibnamefont
  {Stachiotti}}, \bibinfo {author} {\bibfnamefont {M.}~\bibnamefont
  {Sepliarsky}}, \bibinfo {author} {\bibfnamefont {R.~L.}\ \bibnamefont
  {Migoni}},\ and\ \bibinfo {author} {\bibfnamefont {C.~O.}\ \bibnamefont
  {Rodriguez}},\ }\href {https://doi.org/10.1088/0953-8984/11/48/325}
  {\bibfield  {journal} {\bibinfo  {journal} {J. Phys. Condens. Matter.}\
  }\textbf {\bibinfo {volume} {11}},\ \bibinfo {pages} {9679} (\bibinfo {year}
  {1999})}\BibitemShut {NoStop}%
\end{thebibliography}%

\pagebreak
\widetext
\vspace{3pt}
\begin{center}
\textbf{\large Supplementary Material: Ultrafast polarization switching in BaTiO$_3$ by photoactivation of its ferroelectric and central modes}
\end{center}
\setcounter{equation}{0}
\setcounter{figure}{0}
\setcounter{table}{0}
\setcounter{page}{1}
\makeatletter
\renewcommand{\theequation}{S\arabic{equation}}
\renewcommand{\thefigure}{S\arabic{figure}}
\renewcommand{\bibnumfmt}[1]{[S#1]}
\renewcommand{\citenumfont}[1]{S#1}
\renewcommand{\vec}[1]{\ensuremath{\mathbf{#1}}}


\section{Force fields}
The potential energy is a function ${U=U(\{\vec{r}_i\};\boldsymbol{\eta})}$ of
the positions (${\vec{r}_i}$) of all ions, which is parameterized by a material-specific parameter set, ${\boldsymbol{\eta}}$.
The parameter set $\boldsymbol{\eta}$ is determined by a machine-learning process: It is 
fit to the forces, stress tensors, and energy differences calculated by density functional theory (DFT) 
on thermally-disordered crystalline microstructures. Once fit, it is used in molecular dynamics (MD) simulations
to generate a new set of microstructures and the fitting process is repeated to find a new set of parameters.
This process is repeated until self-consistency is reached between the
parameter set that generated the microstructures and the parameter set that is fit to DFT calculations on those microstructures.
The fitting process is described in detail in Ref.~\onlinecite{tangney_jcp_2003}. 

\subsection{Mathematical form of the polarizable ion model with and without variable charges}
The polarizable-ion model with fixed charges is described in detail in Refs.~\onlinecite{sarsam_jcp_2013,tangney_jcp_2003,nemytov_thesis}, 
and its variable-charge (q-Eq) variant is described in detail in Ref.~\onlinecite{nemytov_thesis}.
Here we summarize the mathematical forms of both models.

In both our fixed-charge and q-Eq polarizable-ion models
the total potential energy is the sum, $${U= U^{MS}+ U^{ES} + U^{\text{self}},}$$ of a Morse potential ($U^{MS}$), a 
long-range electrostatic term ($U^{ES}$), and the sum, ${U^\text{self}\equiv \sum_i U_i^{\text{self}}}$, 
of the ions' \emph{self energies}. 

The Morse potential~\cite{Morse_1929,Demiralp_1999} has the form
\begin{equation}
    U^{SR}=\sum_{i,j>i}D_{s_is_j}\left(\exp\left[\gamma_{s_is_j}\left(1-r_{ij}/r_{s_is_j}^0\right)\right]
-2\exp\left[\frac{\gamma_{s_is_j}}{2}\left(1-r_{ij}/r_{s_is_j}^0\right)\right]\right),
\end{equation}
where the sum is over all distinct pairs ${(i,j)}$ of ions  whose
distance of separation, ${r_{ij}\equiv \abs{\vec{r}_i-\vec{r}_j}}$,
is less than a cutoff distance of ${20\;\text{bohr}}$;
${s_i\in\{\text{Ba},\text{Ti},\text{O}\}}$ denotes the atomic species of ion $i$; and 
$D_{s s'}$, $\gamma_{s s'}$ and $r_{s s'}^0$ are parameters that describe
interactions between ions whose species are $s$ and $s'$.

The electrostatic term of both the fixed-charge dipole-polarizable model and the
variable-charge dipole-polarizable model is
the following sum of charge-charge, charge-dipole, and dipole-dipole interactions:
\begin{align*}
    U^{ES} 
=  \frac{1}{4\pi\epsilon_{0}}\sum_{i>j} \Biggl[ q_{i}q_{j}\left(\frac{1}{r_{ij}}-w_{qq}\tilde{I}_{s_is_j}(r_{ij})\right)
&+ \sum_{\alpha}\left(d_i^\alpha q_{j}-q_{i}d_j^\alpha\right)\frac{\partial }{\partial r_{ij}^\alpha}\left(\frac{1}{r_{ij}}-w_{qd}\tilde{I}_{s_is_j}(r_{ij})\right)\\
    &-\sum_{\alpha,\beta}d_i^\alpha d_j^\beta\frac{\partial^2 }{\partial r_{ij}^\beta \partial r_{ij}^\alpha}\left(\frac{1}{r_{ij}}-w_{dd}
\tilde{I}_{s_is_j}(r_{ij})\right)
\Biggr],
\end{align*}
where ${w_{qq}, w_{qd}, w_{dd}\in\boldsymbol{\eta}}$ are parameters of the model;
${\mathbf{d}_i\equiv (d_i^1,d_i^2,d_i^3)}$ is the dipole moment of ion $i$
in Cartesian components; the charge of ion $i$ is ${q_i\equiv q_{s_i}^0+\Delta q_i}$, where
each ${q_{s_i}^0}$ is a parameter of the model and ${\Delta q_i}$ is zero
in the fixed-charge model.
The function
\begin{align*}
\tilde{I}_{s_is_j}(r_{ij}) \equiv c_{s_is_j}e^{-b_{s_is_j}r_{ij}}
\left[\sum_{k=0}^{n_k+1}\frac{(b_{s_is_j}r_{ij})^k}{k!} -\frac{b_{s_is_j}}{n_k+1} 
\sum_{k=0}^{n_k}\frac{(b_{s_is_j}r_{ij})^k}{k!} \right],
\end{align*}
where ${n_k=4}$, and ${b_{ss'},c_{ss'}\in\boldsymbol{\eta}}$ are material-specific parameters. The second
term in this expression is a correction to ${1/r_{ij}}$ that accounts for overlap between ions'
electron clouds.

The self energy, ${U^\text{self}_i(\vec{d}_i,\Delta q_i)}$, of ion $i$ is the energy cost of deforming its spherically-symmetric electron
cloud to give it a dipole moment ${\vec{d}_i}$ and, in the q-Eq model, to change its charge by ${\Delta q_i}$ 
relative to its fixed reference charge, ${q_{s_i}^0}$.
We express it as a Taylor expansion about
the reference state, ${(\Delta q_i,\mathbf{d}_i)=(0,0)}$, as follows:
\begin{equation}
\begin{aligned}
U_i^\text{self}=&U_i^{0}+\frac{\partial U_i^{\text{self}}}{\partial (\Delta q_i)}\Bigg|_{(0,0)}\Delta q_i
+\frac{\partial U_i^{\text{self}}}{\partial (d_i^\alpha)}\Bigg|_{(0,0)}d_i^\alpha
+ \frac{1}{2}\frac{\partial^2U_i^{\text{self}}}{\partial (\Delta q_i)^2}\Bigg|_{(0,0)}\Delta q_i^2\\ 
&+\sum_{\alpha,\beta}\frac{1}{2}\frac{\partial^2U_i^{\text{self}}}{\partial d_i^{\alpha}\partial d_i^{\beta}}\Bigg|_{(0,0)}d_i^{\alpha}d_i^{\beta} 
+\sum_{\alpha}\frac{1}{2}\frac{\partial^2U_i^{\text{self}}}{\partial (\Delta q_i)\partial d_i^{\alpha}}\Bigg|_{(0,0)}\Delta q_id_i^{\alpha}+\cdots\\
=&A_{s_i}\Delta q_i+B_{s_i}\Delta q_i^2+\sum_{\alpha}C_{s_i} (d_i^{\alpha })^2,
\end{aligned}
\end{equation}
where the terms that are linear in ${d_i^\alpha}$ vanish by symmetry and ${A_s, B_s,C_s\in\boldsymbol{\eta}}$ 
are material-specific and ion-specific parameters of the model.

In the fixed-charge polarizable-ion model each polarizable ion has three internal degrees of freedom, namely,
the three components of its dipole moment. In the variable-charge polarizable-ion mode, each ion 
has an additional internal degree of freedom, namely, ${\Delta q_i}$.
The values of the ${\Delta q_i}$'s and the ${\mathbf{d}_i}$'s change
as ions move and are determined at every step of an MD simulation.
Their values at each MD step are those that minimize
${U^{ES}+U^{\text{self}}}$ under the constraint ${\sum_i q_i=0}$. 
In our simulations we found them by iteration to self-consistency, but
they can also be found by direct minimization, matrix inversion, or combinations of these three approaches~\cite{Sarsam2013,nemytov_thesis}.
The isotropic polarizability of species $s$ is ${\alpha_s\equiv 1/2C_s}$.

In the fixed charge polarizable model the charges do not vary (${\Delta q_i=0,\; \forall i}$), 
which is equivalent to $A_s$ and $B_s$ having very large values for all species $s$.

\subsection{Parametrization of the force fields}\label{sec:parametrization}
The set of parameters in the polarizable ion and \textit{q-Eq} model for the 
ground state and photoexcited state with different excitation levels (${x\in\{0,0.05\;\conc, 0.12\;\conc\}}$) 
are determined by minimizing the cost function 
\begin{equation}
    \Gamma_{cost}=w_f \Gamma_{f}+w_{s} \Gamma_{s}+w_{e}\Gamma_{e},
    \label{eqn:weight}
\end{equation}
where $w_{f}$, $w_{s}$ and $w_{e}$ are weights 
and $\Gamma_f$, $\Gamma_s$, and ${\Gamma_e}$ are relative root-mean-squared errors
in forces, stress tensor components, and energy differences, respectively.
These errors are calculated relative to DFT calculations (ground state) and constrained DFT calculations (excited states)
with the PBEsol functional, as described
in Ref.~\onlinecite{Gu2021}.
The force, stress, and energy cost functions are as follows, where the sums over $k$ and $l$ are
sums over different microstructures, the sums over $i$ are sums over ions in a given microstructure, 
and the sums over $\alpha$ and $\beta$ are sums over Cartesian components.
\begin{equation}
\begin{aligned}
    \Gamma_{f}   &=\frac{\sqrt{\sum_{k=1}^n\sum_{i=1}^N\sum_{\alpha=1}^3 
\mid F_{k,i}^\alpha-F_{k,i}^{\alpha,\mathrm{DFT}}\mid^2}}{\sqrt{\sum_{k=1}^n\sum_{i=1}^N\sum_{\alpha=1}^3 \left(F_{k,i}^{\alpha,\mathrm{DFT}}\right)^2}} ,\\ 
   \Gamma_{s}  &=\frac{\sqrt{\sum_{k=1}^n\sum_{\substack{\alpha=1\\\beta\ge\alpha}}^3 \mid \sigma_{\alpha\beta}^{k,\mathrm{FF}}-\sigma_{\alpha\beta}^{k,\mathrm{DFT}}\mid^2}}{B\sqrt{6n}}, \\
     \Gamma_{e}&=\frac{\sqrt{\sum_{\substack{k=1\\ l> k}}^n \mid\left(U_{k}^{\mathrm{FF}}-U_{l}^{\mathrm{FF}}\right) -\left(U_{k}^{\mathrm{DFT}}-U_{l}^{\mathrm{DFT}}\right)\mid^2}}{\sqrt{\sum_{\substack{k=1\\ l> k}}^n \left(U_{k}^{\mathrm{DFT}}-U_{l}^{\mathrm{DFT}}\right)^2}}.
\end{aligned}
\end{equation}
${B}$ is the bulk modulus and $n$ is the number of different microstructures used in the fit.

As mentioned above, several (${\sim 5}$) parameterizations were performed at each value of $x$ to
achieve self-consistency between the microstructures to which the parameters were fit
and the microstructures sampled in MD simulations with the fitted parameters.
Each of these parameterizations involved fitting to ${\sim 20-30}$ microstructures
at temperatures of ${100\;\text{K}}$, ${200\;\text{K}}$, ${300\;\text{K}}$ and ${500\;\text{K}}$.
Each microstructure was a set of atomic positions taken from a long MD simulation, performed
with the previous parameter set, of a ${3\times 3 \times 3}$ supercell (${135}$ atoms).
The (constrained) DFT calculations were performed with the PBEsol functional~\cite{PBE-sol}.

After completing the self-consistent parameterizations, the ability of each parameter
set to fit a set of $20$ microstructures that was not used in the fitting procedure
was tested. The table below reports the values of 
${\Delta F\equiv 100\times \Gamma_f}$,
${\Delta S\equiv 100\times \Gamma_s}$, 
and
${\Delta E\equiv 100\times \Gamma_e}$ at the end of the fitting procedure. The numbers in 
parentheses report how closely they fit the set of microstructures that were not used 
to fit them.
\begin{center}
\vspace{10pt}
\begin{spacing}{1.25}
\begin{adjustbox}{width=0.85\textwidth}
\begin{tabular}{ccccccc}
\hline
 & \multicolumn{3}{c}{\textbf{polarizable ion}} & \multicolumn{3}{c}{\textbf{polarizable ion + q-Eq}} \\ \hline
$x$ 
& ${0}$
& ${0.05\;\conc}$
& ${0.12\;\conc}$
& ${0}$
& ${0.05\;\conc}$
& ${0.12\;\conc}$
\\ \hline
$\Delta F$ (\%) & 13.5 (13.5) & 14.3 (14.4) & 15.2 (14.9) & 12.7 (12.9) & 10.3 (10.5) & 9.8 (9.8)\\ 
$\Delta S$ (\%) & 0.8  (0.8) & 0.7 (0.7) & 0.6 (0.6) & 0.7 (1.1) & 0.6 (0.6) & 0.6 (0.7)\\ 
$\Delta E$ (\%) & 15.7 (16.5) & 13.1 (13.2) & 16.8 (15.9) & 20.8 (20.5) & 9.8 (10.1) & 9.8 (9.9)\\ 
\hline
\end{tabular}
\end{adjustbox}
\end{spacing}
\end{center}
Note that the abilities of the excited-state force fields to fit {\em ab initio} data
is greatly improved by allowing ions to exchange charge.

The parameter sets of the three force fields used for the calculations reported in 
this work are presented in the following three tables.
All parameters are presented in Hartree atomic units, i.e., energies are in hartrees, distances are multiples of the
Bohr radius, and charges are multiples of the magnitude of an electron's charge.
\newpage

\vspace{20pt}
\begin{table}[!ht]
\centering
Parameter set for the ground state polarisable ion (\textit{q-Eq}) force field.
\vspace{10pt}
\begin{tabular}{lccc}
\multicolumn{1}{c}{{[}a.u.{]}} & $w_{qq}=-0.29164027$ & $w_{qd}=0.90656933$ & $w_{dd}=0.66125331$ \\ \hline
\multicolumn{1}{c}{} & O & Ti & Ba \\ \hline
$A_{s_i}$ & -5.5953809072$\times10^{1}$ & -5.4488664280$\times10^{1}$ & -5.4464794819$\times10^{1}$ \\
$B_{s_i}$ & 1.6123648368$\times10^{1}$ & 4.9739200506$\times10^{-1}$ & 4.4100576772$\times10^{1}$ \\
$q_i^{0}$ & -1.313129 & 3.292757 & 0.64663 \\
$\alpha_{s_i}$(=1/2$C_{s_i}$) & 3.618496 & 7.907886 & 8.781469 \\ \hline \hline
 & O-O & O-Ti & O-Ba \\ \hline
$b_{ss'}$ & 1.3281306803$\times10^{0}$ & \begin{tabular}[c]{@{}c@{}}4.1731956696 $\times10^{0}$\end{tabular} & 0 \\
$c_{ss'}$ & 2.3370170775$\times10^{0}$ & \begin{tabular}[c]{@{}c@{}}-2.4783428648 $\times10^{1}$\end{tabular} & 0 \\ \hline
$r_{ss'}^{0}$ & 1.4518159195$\times10^{1}$ & 4.9235055004$\times10^{0}$ & 5.2978447762$\times10^{0}$ \\
$\gamma_{ss'}$ & 1.8239657593$\times10^{1}$ & 1.0768668320$\times10^{1}$ & 8.1405265737$\times10^{0}$ \\
$D_{ss'}$ & 1.5047586448$\times10^{-7}$ & 3.8390122986$\times10^{-3}$ & 1.9949771913$\times10^{-2}$ \\ \hline \hline
 & Ti-Ti & Ti-Ba & Ba-Ba \\ \hline
$b_{ss'}$ & 0 & \begin{tabular}[c]{@{}c@{}}1.5079389253 $\times10^{0}$\end{tabular} & 0 \\
$c_{ss'}$ & 0 & \begin{tabular}[c]{@{}c@{}}-2.3055352480 $\times10^{0}$\end{tabular} & 0 \\ \hline
$r_{ss'}^{0}$ & 9.8996506968$\times10^{0}$ & 1.8090039133$\times10^{1}$ & 1.0528224851$\times10^{1}$ \\
$\gamma_{ss'}$ & 1.9751987630$\times10^{1}$ & 8.2040381263$\times10^{0}$ & 9.6965761677$\times10^{0}$ \\
$D_{ss'}$ & -2.7680295269$\times10^{-5}$ & 1.4853279894$\times10^{-4}$ & 3.5218065048$\times10^{-4}$ \\ \hline
\end{tabular}
\end{table}
\vspace{40pt}

\begin{table}[!ht]
\centering
Parameter set for the low excited state (${x=0.05\;\conc}$) polarisable ion (\textit{q-Eq}) force field.
\vspace{10pt}
\begin{tabular}{lccc}
\multicolumn{1}{c}{{[}a.u.{]}} & $w_{qq}=-0.08884636$ & $w_{qd}=0.87662912$ & $w_{dd}=0.74489846$ \\ \hline 
\multicolumn{1}{c}{} & O & Ti & Ba \\ \hline
$A_{s_i}$ & -5.6179334511$\times10^{1}$ & -5.4275519020$\times10^{1}$ & -4.9571289191$\times10^{1}$ \\
$B_{s_i}$ & 1.1342132824$\times10^{0}$ & 3.3692305688$\times10^{-1}$ & 3.9553553371$\times10^{1}$ \\
$q_i^{0}$ & -1.320255 & 3.064888 & 0.895877 \\
$\alpha_{s_i}$(=1/2$C_{s_i}$) & 3.522399 & 8.137217 & 13.561009 \\ \hline \hline
 & O-O & O-Ti & O-Ba \\ \hline
$b_{ss'}$ & 1.3341615119$\times10^{0}$ & 4.1591996365$\times10^{0}$ & 0 \\
$c_{ss'}$ & 2.2895907601$\times10^{0}$ & -2.4015050922$\times10^{1}$ & 0 \\ \hline
$r_{ss'}^{0}$ & 1.4602328563$\times10^{1}$ & 4.9523852621$\times10^{0}$ & 5.2623630083$\times10^{0}$ \\
$\gamma_{ss'}$ & 1.7888455276$\times10^{1}$ & 1.0874235273$\times10^{1}$ & 8.3757388660$\times10^{0}$ \\
$D_{ss'}$ & 1.5047586448$\times10^{-7}$ & 3.7722549561$\times10^{-3}$ & 1.9125670648$\times10^{-2}$ \\ \hline \hline
 & Ti-Ti & Ti-Ba & Ba-Ba \\ \hline
$b_{ss'}$ & 0 & 1.2300704711$\times10^{0}$ & 0 \\
$c_{ss'}$ & 0 & -1.1446679828$\times10^{0}$ & 0 \\ \hline
$r_{ss'}^{0}$ & 1.1023926673$\times10^{1}$ & 1.4993890190$\times10^{1}$ & 8.8356257981$\times10^{0}$ \\
$\gamma_{ss'}$ & 1.5348843674$\times10^{1}$ & 8.5183077743$\times10^{0}$ & 8.9246248218$\times10^{0}$ \\
$D_{ss'}$ & -5.5127730050$\times10^{-5}$ & 2.1810389044$\times10^{-4}$ & 1.5155781061$\times10^{-3}$ \\ \hline
\end{tabular}
\end{table}

\vspace{20pt}
\begin{table}[!ht]
\centering
Parameter set for the high excited state (${x=0.12\;\conc}$) polarisable ion (\textit{q-Eq}) force field.
\vspace{10pt}
\begin{tabular}{lccc}
\multicolumn{1}{c}{{[}a.u.{]}} & $w_{qq}=0.22401931$ & $w_{qd}=0.87049918$ & $w_{dd}=0.58127161$ \\ \hline
\multicolumn{1}{c}{} & O & Ti & Ba \\ \hline
$A_{s_i}$ & -5.6168920093$\times10^{1}$ & -5.4278802069$\times10^{1}$ & -5.1653085192$\times10^{1}$ \\
$B_{s_i}$ & 5.3373729677$\times10^{-1}$ & 2.9278068326$\times10^{-1}$ & 4.3490046857$\times10^{0}$ \\
$q_i^{0}$ & -1.361335 & 3.114070 & 0.969935 \\
$\alpha_{s_i}$(=1/2$C_{s_i}$) & 3.358510 & 8.495520 & 12.076623 \\ \hline \hline
 & O-O & O-Ti & O-Ba \\ \hline
$b_{ss'}$ & 1.4739956572$\times10^{0}$ & 4.0581913427$\times10^{0}$ & 0 \\
$c_{ss'}$ & 3.1331792230$\times10^{0}$ & -2.3174218450$\times10^{1}$ & 0 \\ \hline
$r_{ss'}^{0}$ & 1.2193833323$\times10^{1}$ & 4.9401679331$\times10^{0}$ & 5.2363447344$\times10^{0}$ \\
$\gamma_{ss'}$ & 1.8427085919$\times10^{1}$ & 1.1177812719$\times10^{1}$ & 8.2993715207$\times10^{0}$ \\
$D_{ss'}$ & 1.5047586448$\times10^{-7}$ & 3.4451652028$\times10^{-3}$ & 1.8349382317$\times10^{-2}$ \\ \hline \hline
 & Ti-Ti & Ti-Ba & Ba-Ba \\ \hline
$b_{ss'}$ & 0 & 1.3069131344$\times10^{0}$ & 0 \\
$c_{ss'}$ & 0 & -1.3001123731$\times10^{0}$ & 0 \\ \hline
$r_{ss'}^{0}$ & 2.7413319056$\times10^{1}$ & 1.4337720195$\times10^{1}$ & 7.9190459031$\times10^{0}$ \\
$\gamma_{ss'}$ & 1.5286377643$\times10^{1}$ & 8.1155034861$\times10^{0}$ & 8.9767757672$\times10^{0}$ \\
$D_{ss'}$ & -1.5124553612$\times10^{-7}$ & 5.0425689333$\times10^{-4}$ & 4.5964618555$\times10^{-3}$ \\ \hline
\end{tabular}
\end{table}
\newpage

\subsection{Accuracy tests of the force fields}\label{sec:quality_tests}

\subsubsection{Lattice structure}\label{subsec:ground_structure}

The calculated stable structures of the ground state rhombohedral (\textit{R3m}) phase and the high-symmetry cubic (\textit{Pm$\bar{3}$m}) are compared with the calculated DFT results and experimental measurements, as presented in Table.~\ref{table:test_groundstruct}.
The structural parameters of these phases remain consistent regardless of temperature or photoexcitation-induced ferroelectric-paraelectric phase transitions.

\begin{table}[!ht]
\centering
\caption{The space- and time- averaged lattice parameter $a$ (\text{\r{A}}) and unit cell angle $\alpha$ were calculated by \textit{NPE} simulations of a 6$\times$6$\times$6 supercell in the ground state, which equilibrates at $T$=100K, and in the excited state with ${x=0.12\;\conc}$.}
\vspace{5pt}
\begin{spacing}{1.25}
\begin{adjustbox}{width=\textwidth}
\label{table:ff_structure}
\begin{tabular}{ccccccc}
\hline
\multirow{2}{*}{} & \multicolumn{1}{c}{\multirow{2}{*}{\begin{tabular}[c]{@{}c@{}}\textbf{ground state}\\ polarizable ion (\textit{q-Eq})\vspace{2pt} \end{tabular}}} & \multicolumn{2}{c}{\textbf{rhombohedral phase} (\textit{R3m})} & \multicolumn{1}{c}{\multirow{2}{*}{\begin{tabular}[c]{@{}c@{}}\textbf{excited state}\\ polarizable ion (\textit{q-Eq})\end{tabular}}} & \multicolumn{2}{c}{ \textbf{cubic phase} (\textit{Pm$\bar{3}$m)}} \\
 & \multicolumn{1}{c}{} & \multicolumn{1}{c}{DFT (PBEsol) {\cite{Gu2021}}} & \multicolumn{1}{c}{exp {\cite{kwei_JPhysChem_1993}}} & \multicolumn{1}{c}{} & \multicolumn{1}{c}{DFT (PBEsol) {\cite{Gu2021}}} & \multicolumn{1}{c}{exp {\cite{hellwege1982ferro}}} \\
\hline
\multicolumn{1}{c}{\begin{tabular}[c]{@{}c@{}}lattice parameter ($\AA$)\end{tabular}} & 4.006 & 4.002 & 4.003 & 3.984 & 3.983 & 4.000 \\
\multicolumn{1}{c}{\begin{tabular}[c]{@{}c@{}}lattice angle ($\alpha$)\end{tabular}} & 89.79 & 89.86 & 89.84 & 89.99 & 90.00 & 90.00 \\
\hline
\end{tabular}
\end{adjustbox}
\end{spacing}
\label{table:test_groundstruct}
\end{table}

\subsubsection{Potential energy surface}\label{subsec:testff_energysurface}
The potential energy landscape in the ground state and photoexcited states provides information on equilibrium atomic positions and the height of the local energy barriers.
These, in turn, indicate the magnitude of the total polarization ($P$) and the coercive field ($E_c$) required for polarization switching.
For the rhombohedral phase ($R3m$) structure, potential energy surfaces are calculated as a function of Ti ion displacement. The Ti ions are step-wisely displaced from the center of the cubic cell structure along the symmetry-lowering direction [111].
Note that the atomic position of Ti is expressed in reduced coordinates as ($\mathrm{\frac{1}{2}}$+$\mathrm{\delta_{Ti}}$, $\mathrm{\frac{1}{2}}$+$\mathrm{\delta_{Ti}}$, $\mathrm{\frac{1}{2}}$+$\mathrm{\delta_{Ti}}$).
Simultaneously, the O atoms are fully relaxed using the steepest descent method.
The local barrier height ($D$) and the equilibrium Ti position displacement ($\delta_{\mathrm{Ti}}^\mathrm{eq}$) are presented in Table.~\ref{table:test_energysurface}.

\begin{table}[!ht]
\caption{The values of the double well-depth, $D$, and the magnitudes of the off-center displacements of 
the Ti atom, $\delta_{\mathrm{Ti}}^\mathrm{eq}$, for the ground state and photoexcited states.
The values of $\delta_{\mathrm{Ti}}^\mathrm{eq}$ are expressed in units of the lattice parameter, $c$.}
\vspace{5 pt}
\centering
\begin{tabular}{cccc}
\hline
\multicolumn{2}{c}{}                                    & \begin{tabular}[c]{@{}c@{}}Well depth ($D$)\\ (meV)\end{tabular} & \begin{tabular}[c]{@{}c@{}}$\delta_{\mathrm{Ti}}^\mathrm{eq}$\\ (in units of c)\end{tabular} \\ \hline
\multicolumn{1}{c}{\multirow{3}{*}{\begin{tabular}[c]{@{}c@{}}\textbf{ground state}\\ ($x$=0)\end{tabular}}}                                                        & \multicolumn{1}{c}{DFT (PBEsol)}                         & 16.64                                                            & $\pm$0.0112                                                                      \\  
\multicolumn{1}{c}{}                                                        & \multicolumn{1}{c}{polarizable ion}     & 13.82                                                                & $\pm$0.0111                                                                                \\ 
\multicolumn{1}{c}{}                                                    & \multicolumn{1}{c}{polarizable ion (\textit{q-Eq})} & 12.22                                                                & $\pm$0.0107                                                                             \\ \hline
\multicolumn{1}{c}{\multirow{3}{*}{\begin{tabular}[c]{@{}c@{}}\textbf{low excited state}\\ (${x=0.05\;\conc}$)\end{tabular}}}  & \multicolumn{1}{c}{DFT (PBEsol)}                         & 8.18                                                                & $\pm$0.0084                                                                                \\ 
\multicolumn{1}{c}{}                                                                                                & \multicolumn{1}{c}{polarizable ion}     & 8.18                                                                & $\pm$0.0098                                                                                \\ 
\multicolumn{1}{c}{}                                                                                                & \multicolumn{1}{c}{polarizable ion (\textit{q-Eq})} & 7.71                                                                & $\pm$0.0089                                                                                \\ \hline
\multicolumn{1}{c}{\multirow{3}{*}{\begin{tabular}[c]{@{}c@{}}\textbf{high excited state}\\ (${x=0.12\;\conc}$)\end{tabular}}} & \multicolumn{1}{c}{DFT (PBEsol)}                         & 0.07                                                                & $\pm$0.0033                                                                                \\ 
\multicolumn{1}{c}{}                                                                                                & \multicolumn{1}{c}{polarizable ion}     & 0.12                                                                & $\pm$0.0013                                                                                \\ 
\multicolumn{1}{c}{}                                                                                                & \multicolumn{1}{c}{polarizable ion (\textit{q-Eq})} & 0.90                                                                & $\pm$0.0040                                                                                \\ \hline
\end{tabular}
\label{table:test_energysurface}
\end{table}

\subsubsection{Phonon dispersions}\label{subsec:testff_phonon}

\begin{figure}[!h]
  \centering
  \includegraphics[width=16.3cm]{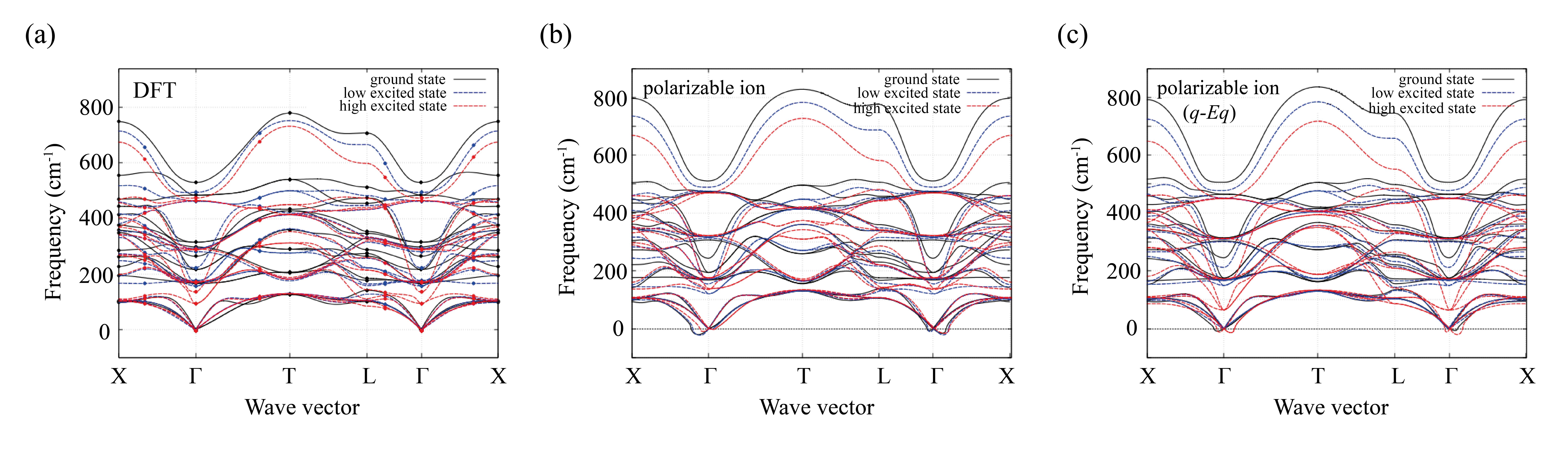}\quad
\caption{Full phonon dispersion curves calculated with (a) constrained DFT~\cite{Gu2021}, (b) polarizable ion model, (c) polarizable ion (\textit{q-Eq}) model. 
The path through high-symmetry points within the Brillouin zone is: X ($\frac{1}{2}$ 0 $-\frac{1}{2}$) $\rightarrow$ $\Gamma$ (0 0 0) $\rightarrow$ T ($\frac{1}{2}$ $\frac{1}{2}$ $\frac{1}{2}$) $\rightarrow$ L (1 $\frac{1}{2}$ 0) $\rightarrow$ $\Gamma$ (1 0 0).}
\label{fig:fftest_phonondispersion}
\end{figure}

The full phonon dispersions of the ground state and excited states are calculated by the polarizable ion force fields at zero temperature (0K), using the finite displacements methods with \textbf{\textit{Phonopy}}~\cite{phonopy}.
Generally, the phonon frequency softens throughout the whole Brillouin zone as the excited carrier density increase (see Fig.~\ref{fig:fftest_phonondispersion}).
Additionally, the softening in the frequency of the FM at the $\Gamma$-point due to photoexcitation is compared to the results of the constrained form of DFT~\cite{Gu2021}, as shown in Fig.~\ref{fig:fftest_softphonon} below.
The polarizable ion model with variable charge (\textit{q-Eq}) agrees well with DFT.
It is noteworthy that applying negative pressure to the unit cell would increase the unit cell's volume, therefore raising the critical excitation level ($x_c$) and adjusting the frequency of the FM phonon.

\begin{figure}[!ht]
\begin{center}
\includegraphics[width=10cm]{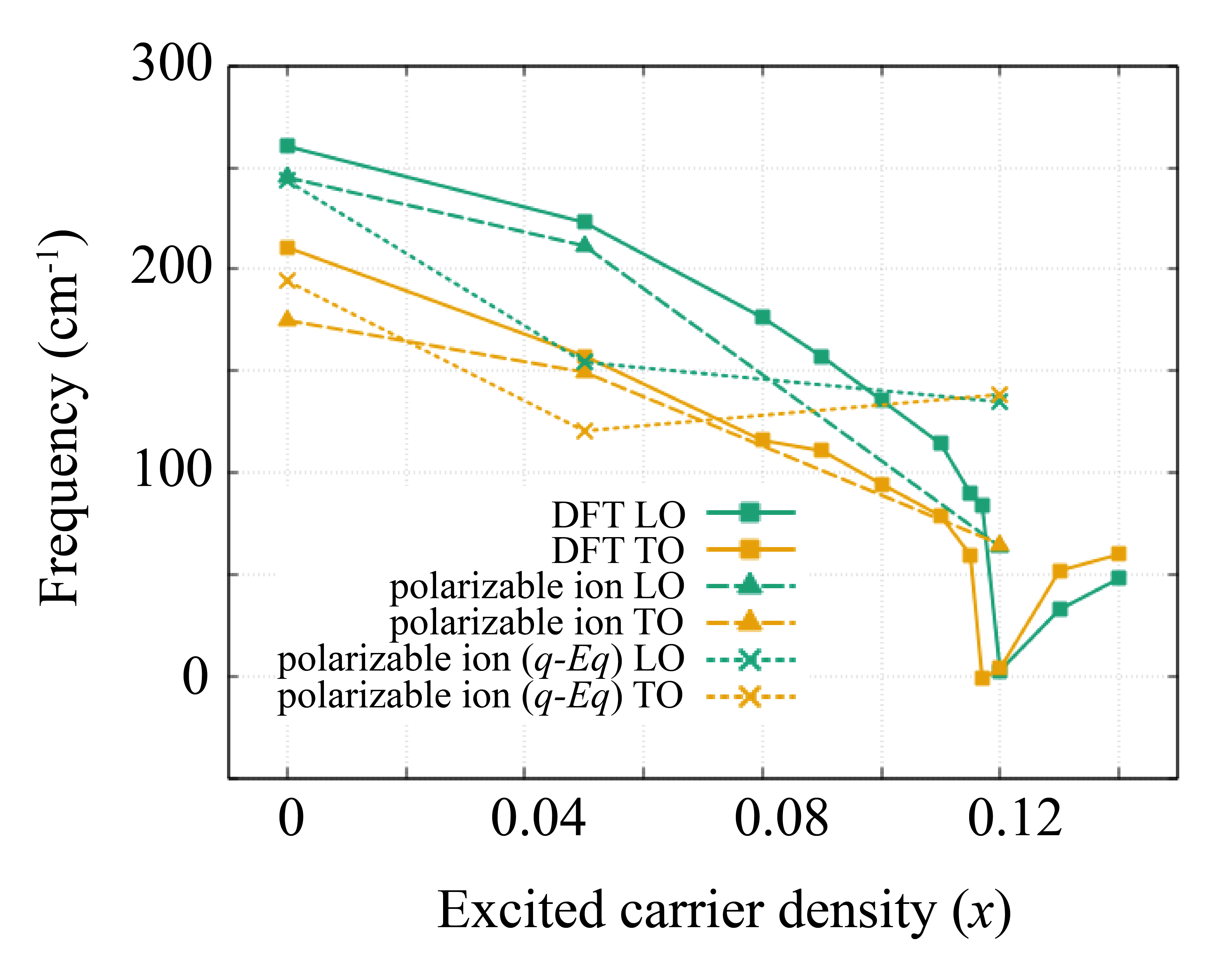}
\end{center}
\caption{The frequency of the FM ($\gpoint$-point) as a function of photoexcited carrier density $x\, (\conc)$.}
\label{fig:fftest_softphonon}
\end{figure}  

\subsubsection{Bulk modulus}\label{subsec:testff_bulkmodulus}

Since ferroelectricity is very sensitive to elastic properties, it is nesscessary to test the polarizable ion force fields for their ability to reproduce these properties.
An important elastic property, the bulk modulus ($B$), measures a material's resistance to compression. We use the Murnagham equation of state~\cite{Murnaghan244} to calculate the bulk modulus, which relates the volume of the cell to the applied pressure. 
\begin{equation}
E_{\mathrm{Murn}}(V)=-E_{0}+\frac{B_{0}V}{B_{0}^{'}}\Big[\frac{(V^{0}/V)^{B_{0}^{'}}}{B_{0}^{'}-1}+1 \Big]-\frac{V^{0}B_{0}}{B_{0}^{'}-1}
\end{equation}
Experiments~\cite{Piskunov2004,kingsmith1994,Tinte_1999} have shown a larger value of the bulk modulus for the cubic phase BaTiO$_3$ ($\sim$170GPa), and $\sim$120GPa for the rhombohedral phase.
These values are in good agreement with the bulk modulus calculated by parametrized force fields, as shown in Table.~\ref{table:fftest_bulkmodulus}.
The ground state exhibits the rhombohedral phase, and with increasing excited carrier density ($x$), the system undergoes a phase transition, entering the high-symmetry cubic phase at $x=0.12 \;\conc$.

\begin{table}[!ht]
\centering
\caption{Bulk moduli calculated by constrained DFT~\cite{Gu2021}, the polarizable-ion force field,
and the variable-charge (q-Eq) polarizable-ion force field.}
\vspace{7pt}
\begin{tabular}{cccc}
\hline
bulk modulus (GPa)                                                           & DFT (PBEsol)  & polarizable ion & polarizable ion + \textit{q-Eq}) \\ \hline
\begin{tabular}[c]{@{}c@{}}\textbf{ground state}\\ (${x=0}$)\end{tabular}      & 120.1  & 116.6          & 126.0                          \\ \hline
\begin{tabular}[c]{@{}c@{}}\textbf{low excited state}\\ (${x=0.05\;\conc}$)\end{tabular}  & 122.4 & 134.6          & 125.7                          \\ \hline
\begin{tabular}[c]{@{}c@{}}\textbf{high excited state}\\ (${x=0.12\;\conc}$)\end{tabular} & 178.1 & 232.3          & 179.8                          \\ \hline
\end{tabular}
\label{table:fftest_bulkmodulus}
\end{table}


\subsubsection{Born effective charges and long-range field corrections}\label{subsec:testff_borncharges}

The Born effective charges, also known as dynamical charges, is a tensor associated with each inequivalent atom in the primitive unit cell.
It measures the change in macroscopic polarization in one direction ($\beta$) caused by a sublattice displacement in a specific direction ($\alpha$). The components of the Born effective charge tensor are defined as: $Z_{\kappa,\beta\alpha}=\Omega\frac{\partial P_{\beta}}{\partial \tau_{\kappa,\alpha}}\Big \vert_{\pes=0}$, where $\pes$ is the external field, and $\Omega$ is the volume of the unit cell. 

In the high-symmetry cubic phase, the three eigenvalues are identical and can be written down as: $\epsilon_{11}=\epsilon_{22}=\epsilon_{33}$, $Z_{\mathrm{Ba},11}=Z_{\mathrm{Ba},22}=Z_{\mathrm{Ba},33}$ and $Z_{\mathrm{Ti},11}=Z_{\mathrm{Ti},22}=Z_{\mathrm{Ti},33}$.
For the O atom, there are two distinct types of eigenvalues: $Z_{\mathrm{O}_{\parallel}}$ and $Z_{\mathrm{O}_{\perp}}$, referring to atomic displacement parallel to the Ti-O bond direction and perpendicular to it, respectively.
In the rhombohedral phase, which generally exists in the ground state, the three chosen primitive vectors are equivalent and rotationally symmetrical with regard to the rhombohedral symmetric axis, [111]. 
Therefore, the three diagonal eigenvalues are also the same. 
The Born effective charges calculated by different force fields for the ground state in the rhombohedral phase are compared in Table.~\ref{table:fftest_bornground} below.
Meanwhile, the magnitudes of the long-range correction (LO-TO splitting at $\gpoint$-point), which scales with $\frac{Z^{*}}{\sqrt{\epsilon_{\infty}}}$, are found to agree well with the \textit{ab-initio} DFPT results, as given in the same Table.~\ref{table:fftest_bornground}.
The excellent agreement between $\frac{Z^{*}}{\sqrt{\epsilon_{\infty}}}$ and the less good agreement between $Z^{*}$ demonstrate that the long-range Coulomb interaction and the corresponding dynamics are accurately captured by force fields. However, whether the force is generated by charges or dipoles is not correctly distinguished.

\begin{table}[!ht]
\centering
\caption{Comparison of Born Effective Charges ($Z$) and the long-range field correction ($\frac{Z}{\sqrt{\epsilon_{\infty}}}$) in the ground state for different models.}
\vspace{5 pt}
\begin{tabular}{cccc}
\hline                                                                                                & DFPT   & polarizable ion & polarizable ion (\textit{q-Eq}) \\ \hline
$Z_{\mathrm{Ba}}$                                                                                       & 2.75  & 1.67            & 1.73                   \\
$Z_{\mathrm{Ti}}$                                                                                       & 6.29  & 4.32            & 4.49                   \\
$Z_{\mathrm{O_{\|}}}$                                                                                   & -4.88 & -4.00           & -3.98                  \\
\begin{tabular}[c]{@{}c@{}}$Z_{\mathrm{O_{\bot} }}$\end{tabular}                                   & -1.94 & -1.00           & -1.12                  \\ \hline
${Z_{\mathrm{Ba}}}/{\sqrt{\epsilon_{\infty}}}$                                                     & 1.12  & 0.94            & 0.98                   \\
\begin{tabular}[c]{@{}c@{}}${Z_{\mathrm{Ti}}}/{\sqrt{\epsilon_{\infty}}}$\end{tabular}       & 2.56  & 2.42            & 2.53                   \\
\begin{tabular}[c]{@{}c@{}}${Z_{\mathrm{O_{\|}}}}/{\sqrt{\epsilon_{\infty}}}$\end{tabular}   & -2.23 & -2.24           & -2.24                  \\
\begin{tabular}[c]{@{}c@{}}${Z_{\mathrm{O_{\bot}}}}/{\sqrt{\epsilon_{\infty}}}$\end{tabular} & -0.78 & -0.56           & -0.63                  \\ \hline
\end{tabular}
\label{table:fftest_bornground}
\end{table}

As shown in Table.~\ref{table:fftest_bornexcited}, the values of Born effective charges decrease with increasing excited carrier density ($x$).
This decrease occurs because the electrons are primarily excited from O (\textit{2p}) state to Ti (\textit{3d}) state, resulting in a reduction in the ion's charge for Ti and its surrounding O atoms.

\begin{table}[!ht]
\centering
\caption{Comparison of Born Effective Charges ($Z$) and the long-range field correction ($\frac{Z}{\sqrt{\epsilon_{\infty}}}$) for the ground state and different photoexcited states using polarizable ion force fields. The excited carriers ($x$) are given in units of $\;\conc$.}
\vspace{5 pt}
\begin{tabular}{cccc}
\hline                                                                                                & \textbf{ground state}   & \begin{tabular}[c]{@{}c@{}}\textbf{low excited state}\\ ($x$=0.05)\end{tabular}  & \begin{tabular}[c]{@{}c@{}}\textbf{high excited state}\\ ($x$=0.12)\end{tabular} \\ \hline
$Z_{\mathrm{Ba}}$                                                                                       & 1.67  & 1.42            & 1.01                  \\
$Z_{\mathrm{Ti}}$                                                                                       & 4.32  & 4.20            & 3.48                   \\
$Z_{\mathrm{O_{\|}}}$                                                                                   & -4.00 & -3.69           & -2.62                  \\
\begin{tabular}[c]{@{}c@{}}$Z_{\mathrm{O_{\bot} }}$\end{tabular}                                   & -1.00 & -0.97           & -0.94                  \\ \hline
${Z_{\mathrm{Ba}}}/{\sqrt{\epsilon_{\infty}}}$                                                     & 0.94  & 0.76            & 0.52                   \\
\begin{tabular}[c]{@{}c@{}}${Z_{\mathrm{Ti}}}/{\sqrt{\epsilon_{\infty}}}$\end{tabular}       & 2.42  & 2.25            & 1.78                   \\
\begin{tabular}[c]{@{}c@{}}${Z_{\mathrm{O_{\|}}}}/{\sqrt{\epsilon_{\infty}}}$\end{tabular}   & -2.24 & -1.98           & -1.34                  \\
\begin{tabular}[c]{@{}c@{}}${Z_{\mathrm{O_{\bot}}}}/{\sqrt{\epsilon_{\infty}}}$\end{tabular} & -0.56 & -0.63           & -0.48                  \\ \hline
\end{tabular}
\label{table:fftest_bornexcited}
\end{table}

\section{\label{sec:infrared_spectra}Infrared absorption spectrum}
The infrared absorption spectrum is obtained by expressing the infrared absorption intensity in terms of the quantum mechanical auto-correlation of the global polarization using perturbation theory.
First, we derive the transition probability from state $| i\rangle$ to state $| f\rangle$ per unit time using {\em Fermi's golden rule}, to the first order of perturbation, which is
\begin{equation}
P_{i \rightarrow f}=\frac{2 \pi}{\hbar}\left|\left\langle f\left|\Delta \cal{H}\right| i\right\rangle\right|^{2} \rho\left(E_{f}\right)
\label{eqn:Pif1}
\end{equation}
where $\Delta \cal{H}$ is the perturbing Hamiltonian, $\rho\left(E_{f}\right)$ is the density of states at the energy $E_{f}$ of the final states, $| i\rangle$ and $| f\rangle$ are the initial and final state, respectively.
Consider a monochromatic electric field of frequency $\omega$,
\begin{equation}  \mathbf{E}_{\omega}=\left|\mathbf{E}_{\omega}\right|\cos(\omega t)\hat{\mathbf{\varepsilon}}
\end{equation}
where $\hat{\mathbf{\varepsilon}}$ is the unit vector along the electric field and $\left|\mathbf{E}_{\omega}\right|$ is the amplitude of the field.
The interaction between the field and the system perturbs the Hamiltonian by:
\begin{equation}
    \Delta \cal{H}=-\mathbf{P}\cdot \mathbf{E}_{\omega}
\end{equation}
where $\mathbf{P}$ is the total polarization operator.
Equation~\ref{eqn:Pif1} then becomes:
\begin{equation}
P_{i \rightarrow f}(\omega)=\frac{\pi \left|\mathbf{E}_{\omega} \right|^2}{2 \hbar^{2}}|\langle f|\hat{\varepsilon} \cdot \mathbf{P}| i\rangle|^{2}\left[\delta\left(\omega_{f i}-\omega\right)+\delta\left(\omega_{f i}+\omega\right)\right]
\label{eqn:Pitof2}
\end{equation}
where $\omega_{f i}=\omega_{f}-\omega_{i}$.
Then, the rate of energy loss from the radiation to the material can be written as:
\begin{equation}
    \begin{aligned}
-\dot{\mathbf{E}}_{\mathrm{rad}} &=\sum_{i} \sum_{f} \rho_{i} \hbar \omega_{f i} P_{i \rightarrow f} \\
&= \frac{\pi \left|\mathbf{E}_{\omega} \right|^2}{2 \hbar} \sum_{f, i} \omega_{f i}\left(\rho_{i}-\rho_{f}\right)\left|\langle f|\hat{\varepsilon} \cdot \mathbf{P}| i\rangle\right|^{2} \delta\left(\omega_{f i}-\omega\right) \\
&= \frac{\pi \left|\mathbf{E}_{\omega} \right|^2}{2 \hbar} \left(1-e^{-\beta \hbar \omega}\right)\omega \sum_{f, i} \rho_{i}\left|\langle f|\hat{\varepsilon} \cdot \mathbf{P}| i\rangle\right|^{2} \delta\left(\omega_{f i}-\omega\right)
    \end{aligned}
    \label{eqn:Pitof3}
\end{equation}
where $\rho_{i}$ is the probability of initially being in state $i$, of an thermal equilibrated initial system as:
\begin{equation}
    \begin{aligned}
        \rho_{f}&=\rho_{i} e^{-\beta \hbar \omega_{\rho i}} \\
        \rho_{i}-\rho_{f}&=\rho_{i}\left(1-e^{-\beta \hbar \omega_{f i}}\right)
    \end{aligned}
\end{equation}
Dividing the average incident energy flux from the radiation that given by the magnitude of the Poynting vector, $\left|\mathbf{S}\right|=\frac{cn}{2\mu_0}\left|\mathbf{E}_{\omega} \right|^2$, we obtain an expression for the absorption coefficient:
\begin{equation}
\begin{aligned}
\alpha(\omega)&=\frac{\mu_0 \pi}{\hbar c n} \left(1-e^{-\beta \hbar \omega}\right)\omega \sum_{f, i} \rho_{i}\left|\langle f|\hat{\varepsilon} \cdot \mathbf{P}| i\rangle\right|^{2} \delta\left(\omega_{f i}-\omega\right) \\
& =A\left(1-e^{-\beta h\omega} \right)\omega\int_{-\infty}^{\infty} dt e^{-i\omega t}\langle \mathbf{P}(t) \cdot \mathbf{P}(0)\rangle  
\end{aligned}
\end{equation}
where $A$ is a constant, and the refractive index, $n$, is assumed to be approximately independent of $\omega$.
This is a quasi-classical expression that incorporates quantum mechanical $\omega$-dependent prefactors into the classical correlation function.
In the expression for the absorption coefficient shown above, we have replaced $\langle \hat{\varepsilon} \cdot \mathbf{P}(t)\ \hat{\varepsilon} \cdot \mathbf{P}(0)\rangle$ with $\langle \mathbf{P}(t)\ \mathbf{P}(0)\rangle$.
This assumption is valid either when the field is parallel to the total polarization density ($\mathbf{E}_{\omega}\parallel \mathbf{P}$), or when the sample is polycrystalline, consisting of many domains with different orientations.


\vspace{9pt}

\end{document}